\documentclass[final]{elsarticle}

\usepackage{lineno,hyperref}
\usepackage{geometry}
\makeatletter
\def\ps@pprintTitle{%
 \let\@oddhead\@empty
 \let\@evenhead\@empty
 \def\@oddfoot{}%
 \let\@evenfoot\@oddfoot}
\makeatother
\makeatletter
\renewcommand{\MaketitleBox}{%
  \resetTitleCounters
  \def\baselinestretch{1}%
  \begin{center}
    \def\baselinestretch{1}%
    \Large \@title \par
    \vskip 18pt
    \normalsize\elsauthors \par
    \vskip 10pt
    \footnotesize \itshape \elsaddress \par
  \end{center}
  \vskip 12pt
}
\makeatother

\usepackage{amsmath}
\usepackage{setspace}
\usepackage{ragged2e}

\usepackage{subcaption}
\usepackage{lmodern,babel,adjustbox,booktabs,multirow}
\begin{document}

\begin{frontmatter}
\title{\textbf{Estimation of thermal load on the nozzle base plate from small plumes at high temperature}}
\author{Kamal Khemani\textsuperscript{1}} 
\author{Pradeep Kumar\textsuperscript{1*}}
\ead[]{pradeepkumar@iitmandi.ac.in}
\author{Ganesh Natarajan\textsuperscript{2}}

\address{\textsuperscript{1} Numerical Experiment Laboratory (Radiation \& Fluid Flow Physics)\\
Indian Institute of Technology Mandi, Himachal Pradesh, 175075, India\\
\vspace{0.1cm}
\textsuperscript{2} Discipline of Mechanical Engineering,\\ Indian Institute of Technology Palakkad, Palakkad, Kerala, 678557, India}
\end{frontmatter}
\doublespacing

\section*{Abstract}
A numerical study is performed to estimate thermal load on the nozzle base plate, which is in the upstream direction to the flow, from three hot plumes of pure ($CO_2$), ($H_2O$) and 50-50 (\%) composition of ($CO_2$) and ($H_2O$) expanding through a convergent-divergent (CD) nozzle in a quiescent medium at 1.1 bar pressure and 298K temperature. The base plate of the nozzle heats up due to thermal radiation, emitting from the hot gases in the form of plumes. The spectral radiative properties of major participating gases such as ($CO_2$), ($H_2O$) are calculated from HITEMP-2010 database. A small CD nozzle which is designed for the perfect expansion of air by 1D calculation with nozzle throat diameter 1.98 mm and area ratio 1.5942, is considered as the design of nozzle for present study \cite{darwish2019simulation}. All three plumes are in the under-expanded state for this CD nozzle and hence expands rapidly at supersonic speed as the plumes exit from the nozzle and forms a series of expansion and compression waves. The hot plumes emanating from the nozzle develop very high temperature in a small vicinity around the base plate, due to diffusion and develop  very high temperature on the base plate. Barring this region, the maximum amount of radiative flux on base plate for these three plumes, i.e., $CO_2$ plume, mixture plume and $H_2O$ plume are 4000 $W/m^2$, 2300 $W/m^2$ and 1300 $W/m^2$, respectively and the maximum temperature developed due to these corresponding fluxes are 323 K, 312 K and 308 K, respectively.
\\\\
\noindent
\textbf{Keywords:} Compressible flow, gas radiation, thermal load, underexpanded\\
 \newpage
 \section*{\textbf{NOMENCLATURE}}
\vspace{0.2cm}
\raggedright{\textbf{English Symbols}}\\
\vspace{0.1cm}
\begin{tabular}{p{0.1\linewidth}p{0.9\linewidth}}
    $c_1,\:c_2$&First and second radiation constants\\
    $c_p$&Specific heat at constant pressure\\
    $e$&Internal energy\\
    $h$&Enthalpy\\
    {$k$}&{Thermal conductivity, turbulent kinetic energy}\\
    {$\hat{n}$}& Unit normal vector\\
    $p$&Pressure\\
	{$q$}&{Heat flux}\\
	{$s$}&{Direction vector}\\
    $t$& Time \\
    $u$&Velocity \\
    $x$&Cartesian coordinate coordinate\\
    $A_r$&Area ratio\\
    {$I_\eta$}&{Spectral intensity}\\
	{$I_{b\eta}$}&{Planck function}\\
    $R$&Universal gas constant\\
    $Y$&Species mass-fraction\\
\end{tabular}
\vspace{0.2cm}
\raggedright{\textbf{Greek Symbols}}\\
\vspace{0.1cm}
\begin{tabular}{p{0.1\linewidth}p{0.9\linewidth}}
	$\beta_\eta$  & Spectral extinction coefficient \\
    $\epsilon$&Emissivity, turbulent dissipation rate\\
    $\eta$ & Wavenumber \\
    $\kappa_\eta$ & Spectral absorption coefficient\\
    $\mu$&Dynamic viscosity\\
    {$\nabla \cdot q$}&{Divergence of radiative heat flux}\\
    {$\Omega$}&{Solid angle}\\
    $\phi$&Azimuthal angle\\
	{$\Phi$}&{Scattering phase function}\\
	$\rho$ & Density of fluid\\
	{$\sigma_{s\eta}$}&{Spectral scattering coefficient}\\
	$\theta$&Polar angle\\
    $\tau$&Viscous stress tensor, transmissivity of gas, optical thickness\\
\end{tabular}
\vspace{0.2cm}
\raggedright{\textbf{Subscript}}\\
\vspace{0.1cm}
\begin{tabular}{p{0.1\linewidth}p{0.9\linewidth}}
	$b$&Blackbody\\
	$c$&Conduction\\
	$cv$&Convection\\
	$eff$&Effective\\
	$\eta$&Spectral\\
	$g$&Gas\\
	$k$&Turbulent kinetic energy\\
	$r$&Radiation\\
	$t$&Turbulent, total\\
	$w$&Wall\\
\end{tabular}

\justifying

\newpage
\section{Introduction}
 The exhaust plume from the nozzle is a product of high temperature and high pressure gases exiting from the combustion chamber. These gases expand rapidly in the convergent divergent (CD) nozzle at supersonic velocities because of the conversion of thermal energy into kinetic energy, which generates the thrust to lift off the rocket. The structure of the plume is non uniform, containing different flow regimes and supersonic shock patterns. It appear as bright luminous flame which emits radiation in the visible, ultraviolet (UV) and infrared (IR) parts of the electromagnetic spectrum \cite{simmons2000rocket}. The major part of plume radiation comes from participating gases like $CO_2$, $CO$ and $H_2O$ which show strong emission of thermal radiation in the infrared region of the spectrum \cite{modest2013radiative}. This heats up the base plate of the rocket and becomes the source of tracking by enemies in the case of missiles, fighter jets and combat aircrafts.\\ 
\indent Tien and Abu-Romia \cite{tien1964method} used analytical method to estimated the amount of radiative heat flux on the rocket base plate from exhaust $CO_2$ and $H_2O$ gas plume with idealised physical models. They evaluated apparent emissivity at base plate from semi infinite cylinder shape for $H_2O$ gas plume for a temperature of $2000^oR$, pressure 1 atm and $CO_2$ gas plume for a temperature of $2500^oR$. Nelson \cite{nelson1992backward} used backward Monte Carlo method to estimate radiative heat flux on rocket base plate from exhaust plume. They further studied the effect of cone angle of exhaust plume and scattering albedo on the base plate heating from plume. The increase in cone angle increased the heat flux on the base plate whereas increase of albedo decreased the heat flux. However, increase in albedo increased the searchlight emission from plume. Baek and Kim \cite{baek1997analysis} calculated the heat load on the base plate from both exhaust plume and searchlight emission from the particles. They used finite volume method to solve radiative transfer equation. Tan et al. \cite{tan2005analysis} conducted a study in which they changed the temperature distribution of plume from isothermal to non-isothermal and concluded that the thermal load on thebase plate reduced 2-3 times for non-isothermal plume. They also observed that by increasing optical thickness of medium the amount of radiative flux on the wall increased. Everson and Nelson \cite{everson1993rocket} developed reverse Monte Carlo method to predict base plate heating from plume due to radiation and found that, reverse Monte Carlo was computationally more efficient than forward Monte Carlo method. This was owing to the fact that only the rays that strikes the target point was only considered. For calculations they used band models for gas spectrum and Henyey-Greenstein function for particle scattering. They performed reverse Monte Carlo calculations for four different cases which included pure scattering plume, gas only emission for main engine plume, solid rocket motor plume and a plume with non-uniform temperature which absorbs, emits and scatters, and finally found that majority of emission is due to alumina particles coming from the centre. While, $H_2O$ and $Al_2O_2$ emitted radiation from the center of the plume and moreover major contribution of emission came from $Al_2O_3$ particles. Kumar and Ramamurthy \cite{sunil} estimated radiative heat load on the rocket base plate using forward Monte-Carlo technique for gray conical plume with axial and radial temperature variations. They found that the radiative heat flux changed drastically with the change in radial temperature profile also the amount of radiative heat flux decreased with the increase in altitude as plume cools down faster. Similar arguments were given by Gu and Baek \cite{gu2019analysis} as they examined radiative heat flux from WSGGM method for a solid rocket motor from which the thermal load was estimated by long plumes of 5 and 10 km. \\
\indent Accurate modelling of heat transfer due to radiation is very necessary for safe and efficient designing of rocket. Estimation of radiative properties of gases is crucial and the most important part in determining heat transfer due to radiation accurately. The radiative properties of participating gases can be calculated using some of the most popular spectral database like High Resolution Transmission Spectroscopic Molecular Absorption database (HITRAN) \cite{rothman2009hitran}, Carbon-Dioxide Spectroscopic Database (CDSD) \cite{tashkun2011cdsd}, High Temperature spectroscopic absorption parameter (HITEMP) \cite{rothman2010hitemp} etc. The spectral absorption coefficients are highly erratic in nature containing millions of spectral lines which attain same value multiple times. This unnecessarily increases the computational cost required to solve the radiation transfer equation (RTE) as the line-by-line method considers calculation for each and every line on the spectrum and is therefore, mostly used only for benchmarking purposes \cite{modest2002full}.\\
\indent Many methods are proposed to reduce the computation resource requirements such as Full spectrum scaled and correlated k-Distribution (FSSK/FSCK) \cite{modest2002full}, Lookup based Full spectrum  K-Distribution \cite{wang2016full}, Spectral line weight sum of gray gases \cite{solovjov2000slw} etc. The accuracy of the above methods is well demonstrated for uniform composition of gases \cite{parvatikar2021benchmark, khemani2023radiative}, however, the variation in composition of gaseous and their mixture poses another level of challenge and further modelling is required \cite{khemani2021radiative}. In order to use look up table based FSK method, some interpolation techniques should be adopted for the properties for current thermodynamic states of gases in the domain. It is evident from the above literature that only a few work is available to calculate the heat load on the rocket base plate, that to with fixed conical plume shape and radiative properties of gases. The general heat transfer applications like, combustion, rocket propulsion, gasification contain numerous thermodynamic states, thus it is useful to generate a database for absorption coefficient at different temperatures, pressures and mole-fractions. The present case is optically thin thus, the RTE is solved using the Planck mean absorption coefficient at different thermodynamic states, from look-up table. The thermal load on the nozzle base plate has been calculated from the accurate solution of flow and temperature fields by solving complete set of governing equation. The radiative property is obtained from the HITEMP-2010 database, stored in the form of lookup table for range of thermodynamic states of gases and utilized during the solution of radiative transfer equation. The thermodynamic states for which data is available can directly be used. Further, the Planck mean absorption coefficient for unavailable thermodynamic states can easily be calculated by using multidimensional linear interpolation technique. The fvDOM numerical method is used for solution of RTE coupled with fluid flow using a pressure based compressible flow application sonicRadFoam, modified from sonicFoam application of OpenFOAM \cite{OpenFoamUserGuide}. Finally it includes the work done due to viscous forces, species transfer equation and RTE with Planck mean absorption-emission model.\\
\indent The manuscript is organised as section 2 describing the problem statement, and section 3 describing the mathematical models and governing differential equations followed by validation in section 4, results and discussions in section 5, and finally the present work is concluded in section 6. \\

\section{Problem description}
\begin{figure}
    \centering
    \includegraphics[width=\textwidth]{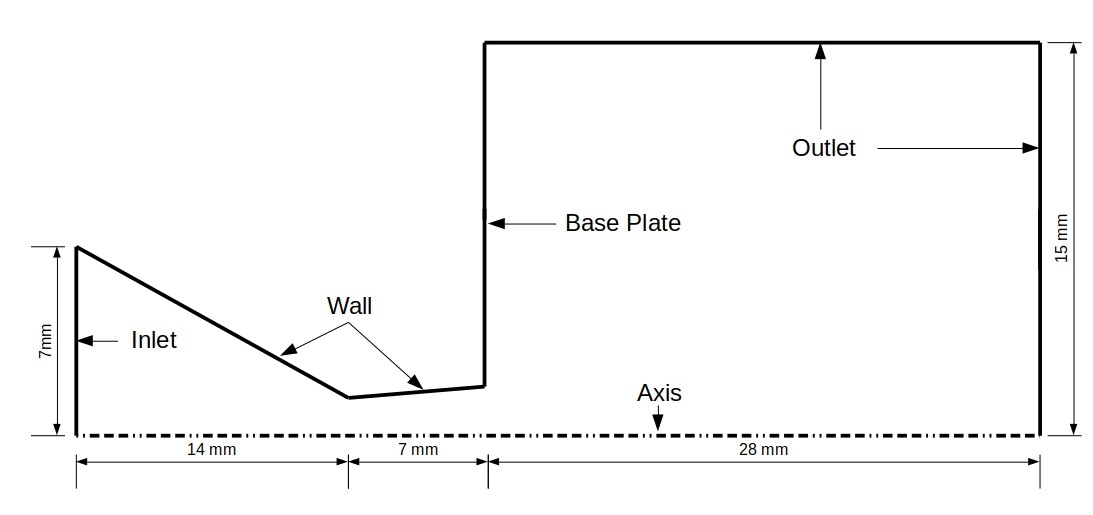}
    \caption{Schematic diagram of geometry for the calculation of the thermal load on the nozzle base plate from the hot plume}
    \label{fig:problDescrip}
\end{figure}

\indent The convergent-divergent (CD) nozzle  has throat diameter and an area-ratio of 1.98 mm and 1.5942, respectively, and the length of convergent and divergent section is 7 mm and 14 mm, respectively as shown in Fig. \ref{fig:problDescrip} which also include the buffer zone for emanating the jet in the atmosphere. The base plate is attached at the end and the fluid expands from a stagnation pressure and temperature of 7.11 bar and 2000 K, respectively, to a quiescent medium at the atmospheric condition of 1 atm pressure and 298K.
The present CD nozzle designed for perfect expansion of air by one dimensional calculation, has been considered for the flow of three plumes whose constituents are pure $CO_2$, pure water vapour and 50-50(\%) $CO_2$ and $H_2O$ from above pressure and temperature. Initially whole domain is filled with $N_2$ gas at 1 atm pressure and 298 K temperature. The following assumptions have been considered for in the present study.

\vspace{0.1cm}
\begin{enumerate}
    \item Reynolds-averaged Navier-Stokes assumption is used to model turbulent flow.
    \item The participating medium only absorbs or emits the thermal radiation but does not scatters.
    \item Refractive index of medium and walls are equal to one.
    \item Turbulence radiation interaction is neglected.
    \item Constant turbulent Prandtl number assumption has been used in the present study:
    \end{enumerate}
\subsection{\textbf{Governing equations}}
\vspace{0.1cm}
The density and temperature fluctuations must be accounted for compressible flow of a fluid along with velocity and pressure fluctuations. To account for these factors, the mass based averaging commonly known as Favre averaging \cite{wilcox1998turbulence,edgarbird}, is used to describe the flow and energy transfer for compressible turbulent fluids.

\noindent which is defined as,

\begin{equation} 
\label{favre}
    {\widetilde{\phi}} = \frac{\overline{\rho \phi}}{\overline{\rho}}
\end{equation}
where, $\rho$ is the density of fluid.
\vspace{0.1cm}
$\phi$ is a scalar and the averaging of density is defined below,
\begin{equation} \label{Timeavgcontinuity}
    \overline{\rho} = \frac{1}{T} \int_{0}^{T}\:{\rho}\:dT
\end{equation}
\vspace{0.1cm}
\begin{equation}  \label{continuity}
    \
    \frac{\partial \overline{\rho} }{\partial t} + \frac{\partial \overline{\rho} \widetilde{u_i}}{\partial x_i}=0
    \
\end{equation}
\vspace{0.1cm}
\begin{equation}   \label{momentum}
   \
    \frac{\partial \overline{\rho} \widetilde{u_i}}{\partial t}+\frac{\partial \overline{\rho} \widetilde{u_i}\widetilde{u_j}}{\partial x_j}=-\frac{\partial \overline{p}}{\partial x_i}+\frac{\partial \widetilde{\tau_{ij}}}{\partial x_j}
    \
\end{equation}
where, 
\begin{equation}
   \widetilde{ \tau_{ij}}=\mu_{eff} \left(\frac{\partial \widetilde{u_i}}{\partial x_j}+\frac{\partial \widetilde{u_j}}{\partial x_i}-\frac{2}{3}\:\delta_{ij}\:\frac{\partial \widetilde{u_k}}{\partial x_k}\right)-\frac{2}{3}\overline{\rho}k\delta_{ij}
\end{equation}
where, $\mu_{eff}$ is the effective dynamic viscosity of fluid which is the summation of molecular and turbulent dynamic viscosity of fluid i.e $(\mu+\mu_t)$ and the molecular viscosity of gases is given by Sutherland
\begin{equation}
   \mu= \frac{A_s\:T^{3/2}}{T+T_s}
\end{equation}
$A_s$ and $T_s$ are Sutherland's constants and depend on the type of gas and it's molecules, and
$\mu_t$ is the turbulent viscosity which is calculated as,
\begin{equation}
    \mu_t=\overline{\rho}\:C_\mu \frac{k^2}{\epsilon}
\end{equation}
where $k$ is turbulent kinetic energy and $\epsilon$ is turbulent dissipation rate and $C_{\mu}$ is the closure constant and these are modelled by two equation $(k.\epsilon)$ turbulence model and given as
\vspace{0.1cm}
\begin{equation}
\frac{\partial\overline{\rho \kappa}}{\partial t} + \frac{\partial \overline{\rho} \widetilde{u_j} \kappa} {\partial x_j} = \frac{\partial}{\partial x_i}  \left[\left (\mu + \frac{\mu_t} {\sigma_\kappa}\right ) \frac{\partial \kappa}{\partial x_i}\right] + P_\kappa - \overline {\rho} \epsilon
\end{equation}
where, $ k = \frac{1}{2} \sum_{i=1}^{3} \frac {\overline{\rho u_i'' u_i''}}{\overline{\rho}} $ is the turbulent kinetic energy, $P_k$ is the production of kinetic energy.
\begin{equation}
 \frac{\partial\overline{\rho }\epsilon}{\partial t} + \frac{\partial \overline{\rho} \widetilde{u_j} \epsilon} {\partial x_j} = \frac{\partial}{\partial x_i}  \left[\left (\mu + \frac{\mu_t} {\sigma_\epsilon
 }\right ) \frac{\partial \epsilon}{\partial x_i}\right] + C_{\epsilon1} \frac{\epsilon}{\kappa} P_\kappa - C_{\epsilon2} \overline{\rho} \frac{\epsilon^2}{\kappa} P_\kappa
 \end{equation}
 where, $ \epsilon = \nu \frac {\widetilde{\partial u_i''\partial u_i''}}{\partial x_j x_j} $ is the turbulent disspation rate and the value of closure constants are as below.
 $C_\mu$ = 0.09, $\sigma_k$ = 1, $sigma_\epsilon$ = 1.3, $C_{\epsilon1}$ = 1.44, $C_{2}$ = 1.92
\vspace{0.1cm}
The pressure is calculated from equation of state for ideal gas law as,
\begin{equation}  \label{eqofstate}
    \
    \overline{p}=\overline{\rho}R\widetilde{T}
    \
\end{equation}
where, $R$ is universal gas constant and $T$ is temperature.
\vspace{0.1cm}
The distribution of species is calculated by species transport equation as below
\begin{equation}  \label{specietransport}
       \frac{\partial \overline{\rho_i} \widetilde{Y_i}}{\partial t}+\frac{\partial \overline{\rho_i}\widetilde{u_i}\widetilde{Y_i}}{\partial x_i}=\frac{\partial}{\partial {x_i}} \left(-\rho \mu_{eff}\frac{\partial \widetilde{Y_i}}{\partial {x_i}}\right)
\end{equation}
where, $Y_i$ is species mass-fraction and is given as, \\
\begin{equation}
    Y_i=\frac{\overline{\rho_i}}{\overline{\rho}}
\end{equation}
\vspace{0.1cm}
The distribution of temperature field is calculated from the energy equation as below
\begin{equation}   \label{energy}
    \frac{\partial \overline{\rho} \widetilde{E}}{\partial t}+\frac{\partial \overline{\rho} \widetilde{u_j}\widetilde{E}}{\partial x_j}+\frac{\partial \widetilde{u_j} \overline{p}}{\partial x_j}=-\frac{\partial \widetilde{q_j}}{\partial x_j}+\frac{\partial \widetilde{u_j} \widetilde{\tau_{ij}}}{\partial x_j}
\end{equation}
where, $E$ is the total energy which includes internal energy $e$, kinetic energy $K$ and turbulent kinetic energy $k$. The heat flux is defined as,
\begin{equation}
    \overline{q_j}=-\frac{c_p\mu_{eff}}{Pr}\frac{\partial \overline{T}}{\partial x_i}+\widetilde{q_r}
\end{equation}
 $c_p$ depends on temperature and are taken from JANAF table of thermodynamics and given as below,
\begin{equation}
    c_p=R((((a_4T+a_3)T+a_2)T+a_1)T+a_0)
\end{equation}
$a_0,\:a_1,\:a_2,\:a_3,\:a_4$ are constants of polynomial,
\begin{equation}
\
q_r=\int_{0}^{\infty}\int_{4\pi}I_{\eta}(\hat{s})\:|\hat{n} \cdot \hat{s}|\:d\Omega\: d\eta
\
\end{equation}
where $q_r$ is the radiative heat flux which can be calculated on the wall, $\hat{n}$ is the surface normal vector, $\partial q_r$/$\partial x_j$ is the divergence of radiative heat flux and can be calculated as,
\begin{equation*} \label{divRTE}
    \nabla \cdot q = \int_0^\infty \kappa_\eta\left(4\pi I_{b\eta}-\int_{4\pi}I_\eta \: d\eta \right) d\eta 
\end{equation*}
or
\begin{equation} \label{divRTEG}
    \nabla \cdot q = \int_0^\infty \kappa_\eta\left(4\pi I_{b\eta}-G_\eta \right) d\eta 
\end{equation}
where $\eta$ is the wavenumber, $I_{b\eta}$ is the Planck function and $\kappa_\eta$ is the spectral absorption coefficient, $G_\eta$ is spectral irradiation, $I_\eta(\hat{s})$ is the intensity field which is obtained by solving the radiative transfer equation (RTE) as explained in the subsequent paragraph.
The above equations are subject to boundary conditions as given in table \ref{tab:bcplume}.

\begin{table}[!b]
\centering
\caption{Boundary conditions for plume with thermal radiation simulation}
\label{tab:bcplume}
\begin{adjustbox}{width=\textwidth}
\begin{tabular}{@{} l*3l @{}}
\toprule
\textbf{Fields} &
  \textbf{Inlet} &
  \textbf{Outlet} &
  \textbf{Wall} \\ \midrule
Pressure ($p$) &
  \begin{tabular}[c]{@{}l@{}}totalPressure\\ $P_o=P+0.5 \: \rho\: U^2$\\ $P_o$ = 7.11 bar\end{tabular} &
  \begin{tabular}[c]{@{}l@{}}fixedValue\\ $P$=1 atm\end{tabular} &
  \begin{tabular}[c]{@{}l@{}}zeroGradient \\ $\nabla P=0$\end{tabular} \\ \\
Velocity ($U$) &
  \begin{tabular}[c]{@{}l@{}}pressureInletOutletVelocity\\ $P_o=P+0.5\: \rho \: U^2$\\ inflow: U = (0,0,0)\\ outflow: $\nabla U=0$\end{tabular} &
  \begin{tabular}[c]{@{}l@{}}inletOutlet\\ inflow: U = (0,0,0)\\ outflow: $\nabla U=0$\end{tabular} &
  \begin{tabular}[c]{@{}l@{}}noSlip\\ U = (0,0,0)\end{tabular} \\  \\
Temperature ($T$) &
  fixedValue $T= 2000 \: K$ &
  \begin{tabular}[c]{@{}l@{}}zeroGradient \\ $\nabla T=0$ \end{tabular} &
  \begin{tabular}[c]{@{}l@{}} $q_c+q_r=0$ \cite{chanakya2022investigation} \end{tabular} \\ \\
  \begin{tabular}[c]{@{}l@{}}Species (x) \end{tabular} &
  \begin{tabular}[c]{@{}l@{}}fixedValue $x = 1$ \\ for pure $H_2O$ plume \end{tabular} &
  \begin{tabular}[c]{@{}l@{}}zeroGradient \\ $\nabla x=0$ \end{tabular} &
  \begin{tabular}[c]{@{}l@{}}zeroGradient \\ $\nabla x=0$ \end{tabular} \\ \bottomrule
\end{tabular}%
\end{adjustbox}
\end{table}
\vspace{0.1cm}
The intensity field in equation \ref{divRTEG} is obtained by solving the spectral radiative transfer equation (s-RTE) for absorbing emitting (not scattering) medium as,
\begin{equation}\label{absoemitRTE}
\
\frac{dI_{\eta}}{ds}=\kappa_{\eta}I_{b\eta}-\kappa_{\eta}I_{\eta}
\
\end{equation}
the above equation is subjected to boundary condition,
\begin{equation} \label{bcrte}
     I_\eta(r_w,\hat{s})=\epsilon_{w_\eta}I_{b\eta}(r_w)+\frac{1-\epsilon_{w\eta}}{\pi}\int_{\hat{n}\cdot\hat{s}>0}I_\eta(r_w,\hat{s})\:|\hat{n}\cdot\hat{s}|\: d\Omega \quad (\hat{n}\cdot \hat{s}<0)  
\end{equation}
where, $\epsilon_{w\eta}$ is the spectral wall emissivity, $I_\eta$ is the spectral intensity along $\hat{s_{i}}$, $I_{b\eta}$ is the Planck function, $\kappa_{\eta}$ is the spectral absorption coefficient, $\eta$ is the wavenumber, and $\Omega$ is the solid angle.
\vspace{0.1cm}
The length scale of the current problem is very small, i.e., the optical length $\tau=\kappa_\eta L<< 1$, this means that the absorptivity of the medium is far less than 1, therefore, the most of the radiation energy will escape the medium without getting absorbed. Thus, the radiative source term (Eq. \ref{divRTEG})
\begin{equation*}
\int_0^\infty \kappa_\eta4\pi I_{b\eta}d\eta << \int_0^\infty \kappa_\eta G_\eta d\eta 
\end{equation*}
The radiative source term Eq. \ref{divRTEG} becomes
\begin{equation*} \label{divRTEGn}
    \nabla \cdot q = \frac{\int_0^\infty \kappa_\eta 4\pi I_{b\eta} d\eta }{\int_0^\infty I_{b\eta} d\eta} \int_0^\infty I_{b\eta} d\eta\\
    = 4\kappa_p\sigma T^4
\end{equation*}
where, $\kappa_p$ is the Planck mean absorption coefficient. Therefore, the solution for the present case can be obtained by Planck Mean absorption coefficient based radiation property model. Thus, the RTE becomes,
\begin{equation} \label{eq:}
\frac{dI_p}{ds} = \kappa_p \cdot(I_b - I_p) \:,
\end{equation}
with boundary conditions,
\begin{equation}
    I_p = \epsilon_w I_b + \frac{1-\epsilon_w}{\pi} \int_{\hat{n}\cdot\hat{s}>0}I_p\:|\hat{n}\cdot\hat{s}|\: d\Omega \quad (\hat{n}\cdot \hat{s}<0)  
\end{equation}
\indent The Planck mean absorption coefficients are calculated for the range of thermodynamic states of gases in the certain intervals as mentioned in (\cite{khemani2023radiative}) and stored in the form of lookup table. Furthermore, interpolation techniques are employed to calculate the absorption coefficient which are not available in the lookup table.\\

The radiative heat transfer, work done due to viscous forces and species transport models have been added into the existing application "sonicFOAM" of the OpenFOAM and named as "radSonicFOAM". The algorithms of the new application is described below and has been extensively verified and validated as explained in the subsequent section and finally, has been used for the estimating the thermal load on the nozzle base plate.
\subsection{\textbf{Numerical Procedure and solution algorithm for solving plume flow with radiation}}
\begin{figure}
    \centering
    \includegraphics[height=1.2\textwidth, width=\textwidth]{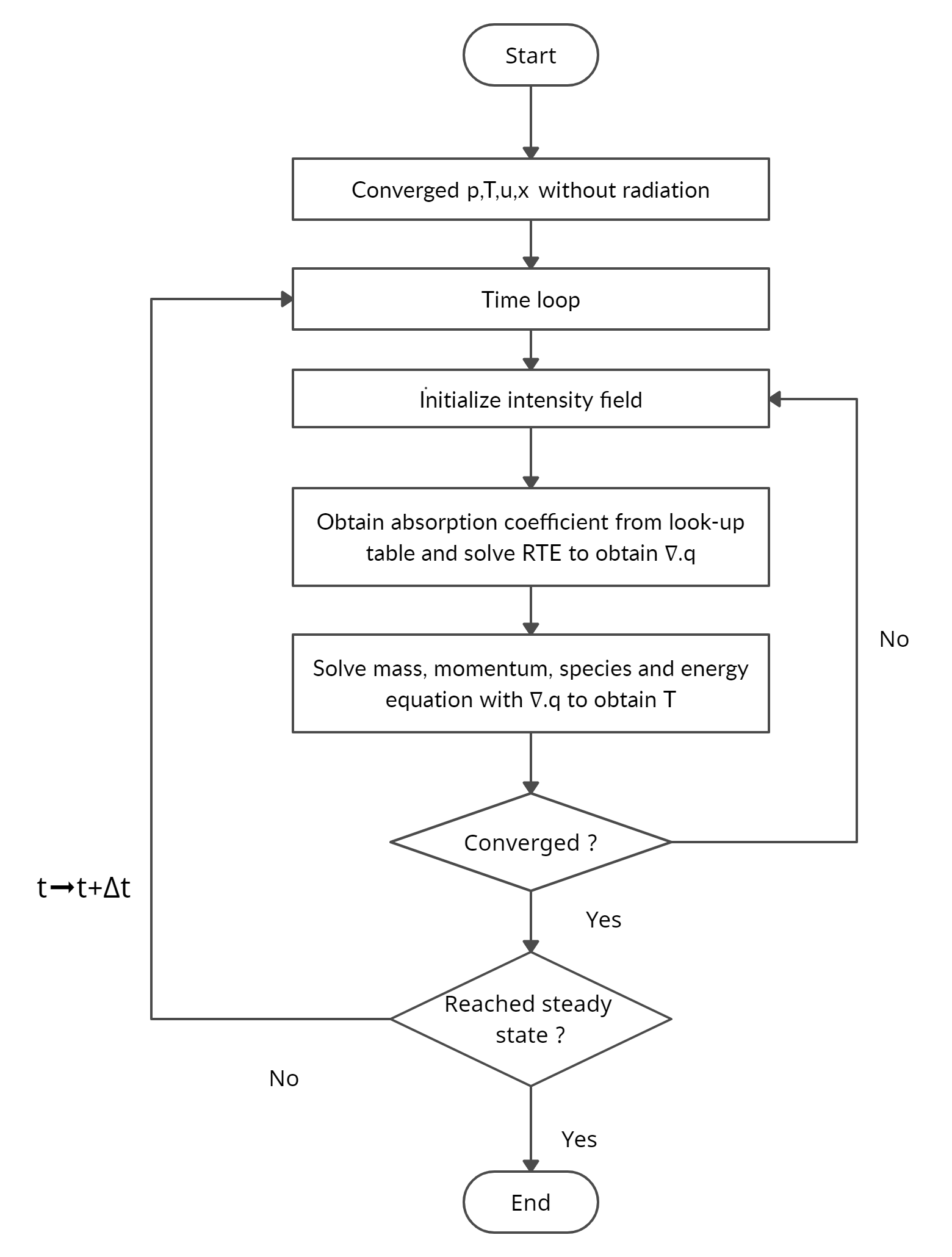}
    \caption{Flow chart for the solution of high temperature and pressure plume flow with radiation}
    \label{fig:plume algo}
\end{figure}

\vspace{0.1cm}
The above mass, momentum, species, energy and radiation transfer equation are discretized using finite volume method \cite{Pradeepthesis}. Further second order upwind scheme is used for the face value interpolation and final set of algebraic equation is solved iteratively, by the SIMPLE algorithm till the residual for mass, momentum, species, energy and radiation reaches to $10^{-5}$ level. The algorithm of above solution method is stated below,
\begin{itemize}
    \item[1.] Initialize pressure, velocity, species and temperature field.
    \item[2.] Solve mass, momentum, species transport and energy equations without radiation till convergence.
    \item[3.] Using converged field, initialize intensity field.
    \item[4.] Calculate Planck mean absorption coefficient from the converged field of temperature, pressure and mole-fraction of species using the Planck mean look-up table and solve RTE till convergence.
    \item[5.] Compute divergence of radiative heat flux.
    \item[6.] Update the temperature field with radiation sink term.
    \item[7.] Repeat 2 to 6 until all the fields reach at steady state.
    furthermore, the flow diagram of the above algorithm is shown in fig \ref{fig:plume algo}.
\end{itemize}
\section{\textbf{Verification and validation studies}}
 The above mathematical modelling and solution algorithm are verified in three steps
 \begin{itemize}
    \item The calculated radiative properties are verified.
    \item The incompressible flow solution is verified with the published result.
    \item The radiative heat flux on the base plate is verified from the assumed shape of the plume in the sections below.
    \end{itemize}
\subsection{\textbf{Verification of Planck mean absorption coefficient of pure \texorpdfstring{$H_2O$ and $CO_2$}{lg}}}
\begin{figure}
    \centering
    \includegraphics[width=0.6\textwidth,height=0.6\textwidth]{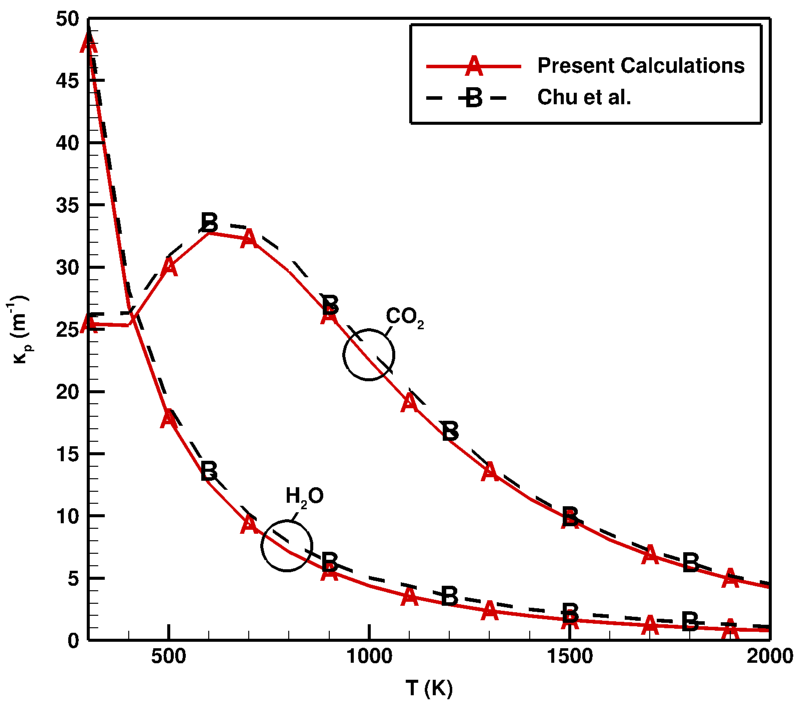}
    \caption{Variation of Planck mean absorption coefficient of pure $H_2O$ and $CO_2$ with different temperature at 1 bar pressure}
    \label{fig:planckmean}
\end{figure}

\vspace{0.1cm}
The Planck mean absorption coefficients obtained for $H_2O$ and $CO_2$ for various temperatures from HITEMP-2010 using in-house C++ code \cite{bartwal2017calculationH2O,bartwal2017calculationCO2,bartwal2018calculation}, match with good agreement from Chu et al. \cite{chu2014calculations}as in Figure \ref{fig:planckmean}.
The Planck mean absorption coefficient of $H_2O$ decreases exponentially with increase in temperature, whereas it first increases up to a temperature of 750 K then decreases till 2000 K for $CO_2$. The Planck mean absorption coefficient of $H_2O$ is higher than $CO_2$ at lower temperatures, however, this is opposite for higher temperature. This difference, decreases with increase in temperatures of compressible flow. 
\vspace{0.1cm}
\subsection{\textbf{Validation of compressible flow field}}
\begin{figure}
    \centering
    \begin{minipage}{0.45\textwidth}
        \centering
        \includegraphics[width=0.9\textwidth]{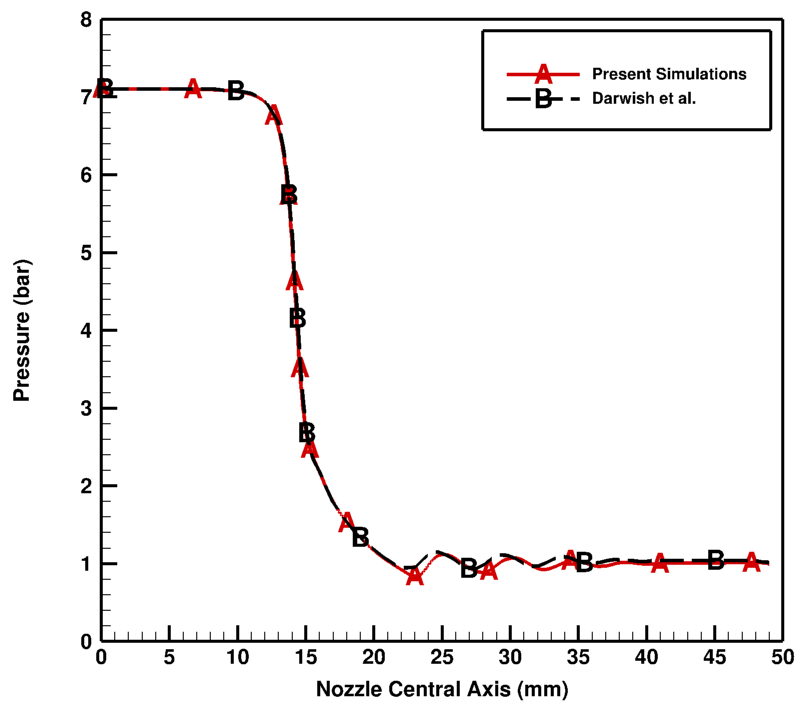}
    \caption{Variation of pressure along the axis of geometry}
    \label{fig:p_air}
    \end{minipage}\hfill
    \begin{minipage}{0.45\textwidth}
        \centering
        \includegraphics[width=0.9\textwidth]{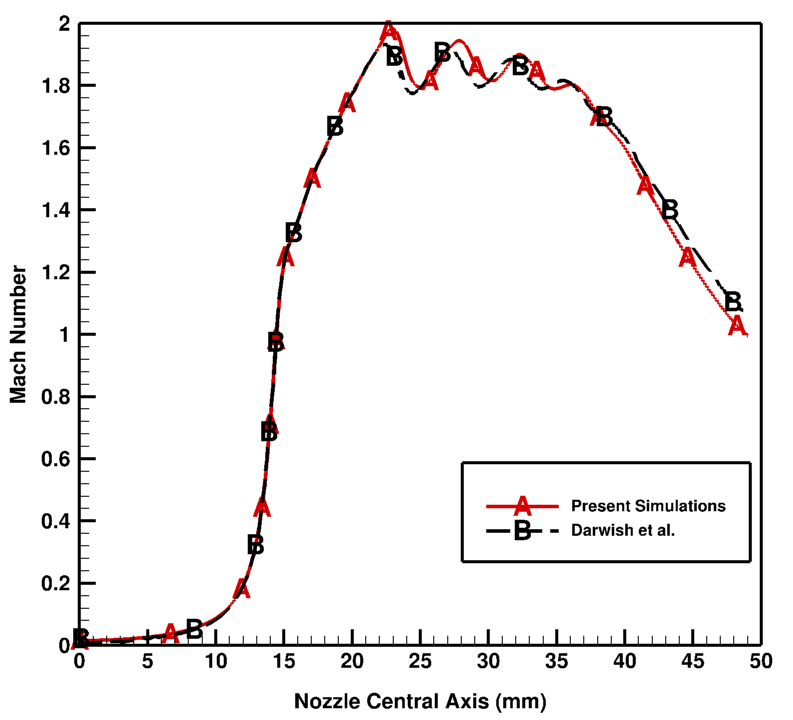}
    \caption{Variation of Mach number along the axis of geometry}
    \label{fig:Ma_air}
    \end{minipage}
\end{figure}

\vspace{0.1cm}
Darwish et al. \cite{darwish2019simulation} have designed a convergent divergent (C-D) nozzle using one dimensional flow isentropic relations for perfect expansion conditions for air. The designed C-D nozzle has an exit diameter of 2.5 mm and throat diameter of 1.98 mm, thus the area ratio $A_r=1.5942$. The schematic diagram of C-D nozzle with buffer section where flow eminates is shown in Fig. \ref{fig:problDescrip}. They simulated the flow using OpenFOAM for axisymmetric geometry for this nozzle along with the buffer zone. They further performed experiments to visualize the flow using shadow-graphic technique. In the present study, we will be using the same nozzle to validate pressure based compressible flow application "sonicFOAM".
The air is allowed to expand from 7.1 atm pressure and 288 K to a quiescent medium at 1 atm pressure. The boundary conditions used for this case is same as given in Table 1 except the temperature at the inlet is 288 K and the walls are at zeroGradient ($\nabla \cdot T = 0$) boundary condition for temperature.
The flow is simulated for axisymmetric geometry by creating a wedge of angle $\theta=2.5^o$ of unit cell in $\theta$ direction. It contains 38,400 cells and the distance of first cell center from the wall is maintained at $y^+ \approx 30$. The standard $k-\epsilon$ model has been used to model turbulence. Pressure-implicit split algorithm (PISO) is used to solve the governing flow and energy equations. Thermophysical and transport properties for air are taken constant as, $C_p=1005\: kJ/kgK$, $\gamma=1.4$, $\mu=1.789\times10^{-5}PaS$ and $Pr=0.7$. The time step used for the present simulation is $10^{-8}\: s$.
The simulation has been performed for $7\:ms$. The pressure and Mach number variation along centerline of nozzle along with the results reported by Darwish et al. \cite{darwish2019simulation}, are plotted in Figure. \ref{fig:p_air} and \ref{fig:Ma_air}, respectively. The present results are in good agreement with the literature results . There are no shocks or sudden discontinuities inside the nozzle as the flow is perfectly expanded inside the nozzle. Since, the nozzle is designed with 1D isentropic calculations and the present simulations are performed for 2D axisymmetric case, there is deviation from 1D isentropic flow. Thus the small expansion and compression waves are formed which create small diamond pattern that can be seen in profiles of pressure and Mach number along the axis of geometry.

\subsection{\textbf{Verification of Rocket base plate heating with assumed plume shape}}
\begin{figure}
    \centering
    \begin{minipage}{0.45\textwidth}
        \centering
        \includegraphics[width=0.9\textwidth]{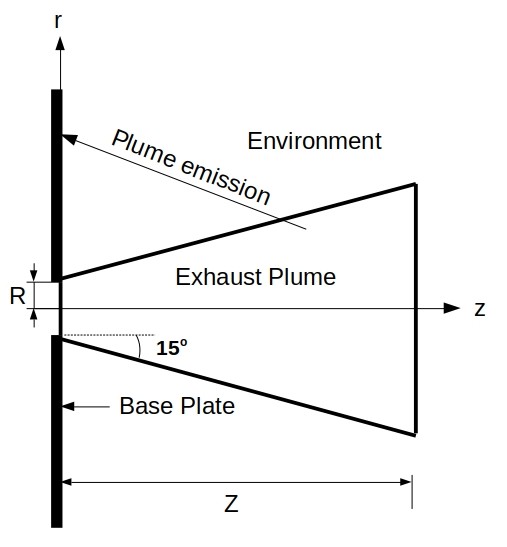}
    \caption{Geometry of conical plume}
    \label{fig:baekplumegeo}
    \end{minipage}\hfill
    \begin{minipage}{0.45\textwidth}
        \centering
        \includegraphics[width=0.9\textwidth]{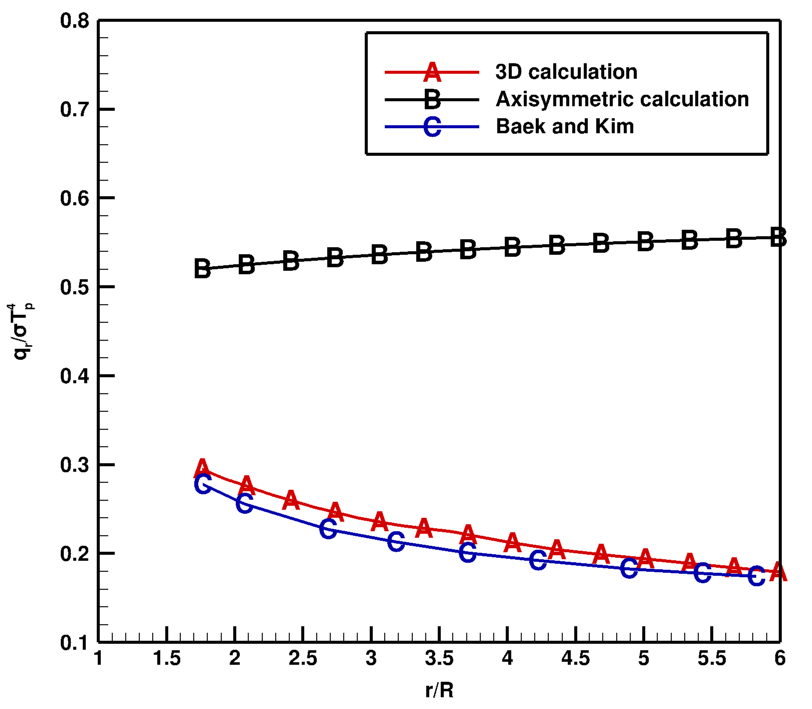}
    \caption{Variation of non-dimensional radiative heat flux by axisymmetric and 3D RTE solution at the base plate from assumed plume shape}
    \label{fig:fluxplumeheating}
    \end{minipage}
\end{figure}

\vspace{0.1cm}
\indent The axisymmetric approximation for RTE has been tested for rocket base plate heating problem from fixed plume shape. The plume is assumed as connical shape with half cone angle of $15^o$ having non-dimensional length $Z/R=50$ as shown in Figure. \ref{fig:baekplumegeo}. The temperature of the plume $T_p$ is uniform. The environment is assumed to be cold and non-participating i.e., $\kappa=0$ and the absorption coefficient of plume is $\kappa=0.5\:m^{-1}$.

\begin{figure}
\begin{subfigure}{.5\textwidth}
  \centering
    \includegraphics[width=\textwidth]{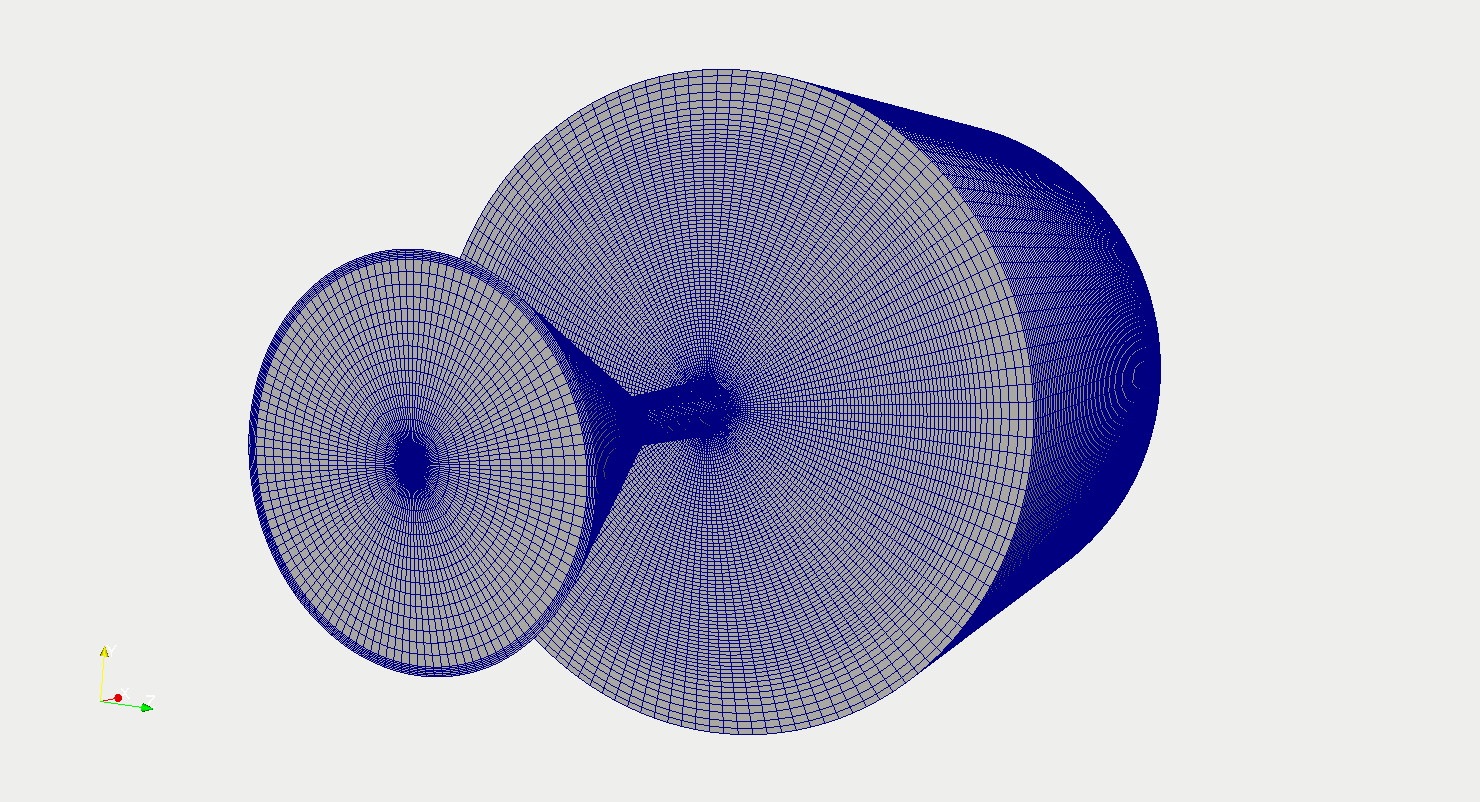}
    \caption{}
    \label{fig:mesh1}
    \end{subfigure}
   \begin{subfigure}{.5\textwidth}
  \centering
  \includegraphics[width=\textwidth]{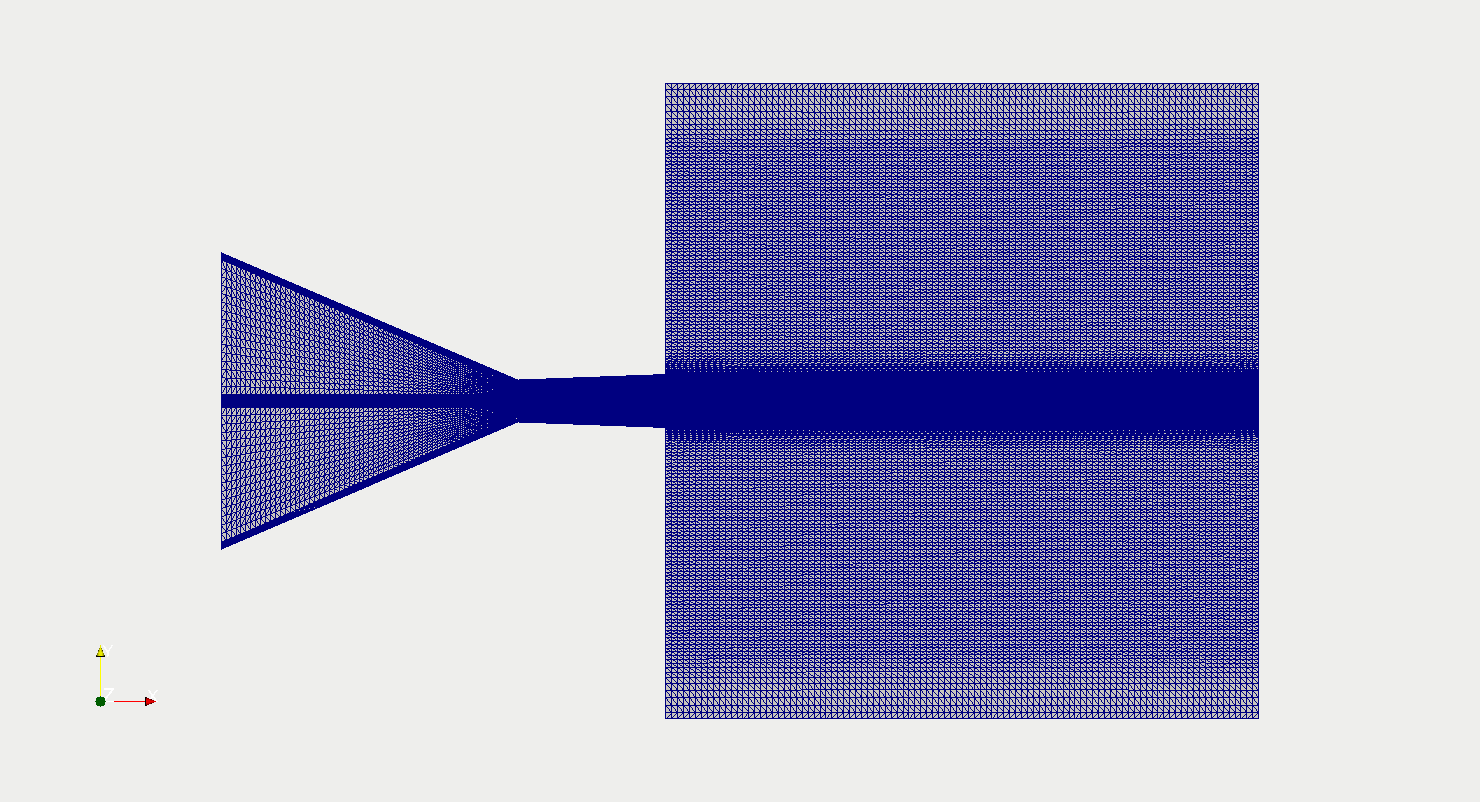}
    \caption{}
    \label{fig:mesh2}
      \end{subfigure}
      \caption{(a) Three dimensional geometry and meshing for simulation of plumes with radiation; (b)Cross sectional view of three dimensional geometry}
\end{figure}

Figure \ref{fig:fluxplumeheating}, shows the radiative heat flux at the base plate from exhaust plume by both axisymmetric and three-dimensional calculations. The result obtained from 3D simulations is in good agreement with the results published by Baek and Kim \cite{baek1997analysis}, whereas axisymmetric simulation result of radiative transfer equations is very far from the result published. This requires reformulation of axisymmetric approximation of radiative heat transfer in OpenFOAM. Therefore, a three dimensional geometry has been used for the further simulations as shown in Figure. \ref{fig:mesh1}.

\section{Results and discussion}

\indent The heating of rocket base plate by thermal radiation from different plumes made of constituents of pure $H_2O$ plume, $CO_2$ plume and 50\%- 50\% mixture of $H_2O$ and $CO_2$ plume are studied numerically with OpenFOAM, an open source CFD package. The present simulations are carried out on a full 3D geometry with a pressure based compressible flow application sonicRadFoam. It has additional features than existing sonicFoam, like work done due to viscous forces in energy equation, species transport equation and emission/absorption due to gaseous radiation. The Planck mean radiation heat transfer model with multidimensional linear interpolation technique for properties is also incorporated to perform radiation heat transfer calculations due to validity of optically thin approximation.

The results of thermal load on the rocket base plate from exhaust plume of three different constituents, i.e., pure $H_2O$ plume, pure $CO_2$ plume and 50\%-50\% mixture of $H_2O$ and $CO_2$ plume are presented in the subsequent sections.

\subsection{Pure $H_2O$ plume}
\begin{figure}

\begin{subfigure}{.5\textwidth}
  \centering
  \includegraphics[width=.7\linewidth,height=.7\linewidth]{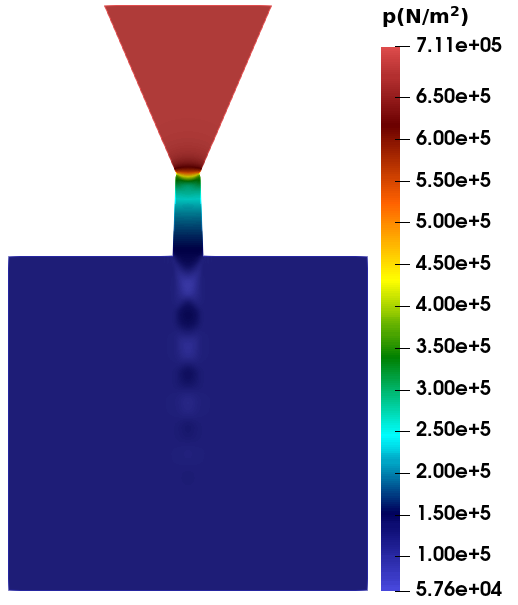}  
  \caption{}
  \label{fig:pconth}
\end{subfigure}
\begin{subfigure}{.5\textwidth}
  \centering
  \includegraphics[width=.7\linewidth,height=.7\linewidth]{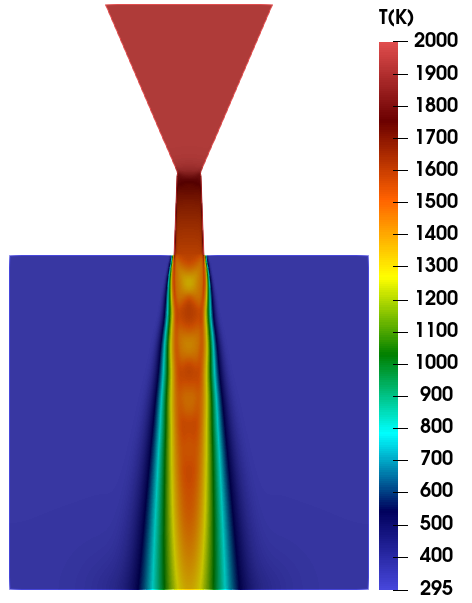}  
  \caption{}
  \label{fig:tconth}
\end{subfigure}

\vspace{1cm}

\begin{subfigure}{.5\textwidth}
  \centering
  \includegraphics[width=.8\linewidth,height=.7\linewidth]{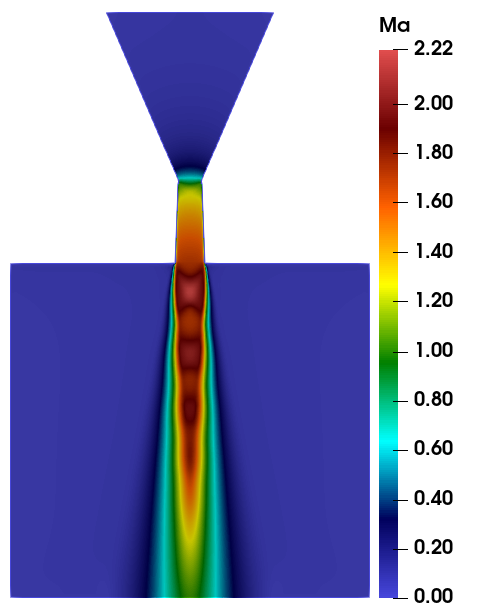}  
  \caption{}
  \label{fig:maconth}
\end{subfigure}
\begin{subfigure}{.5\textwidth}
  \centering
  \includegraphics[width=.8\linewidth,height=.7\linewidth]{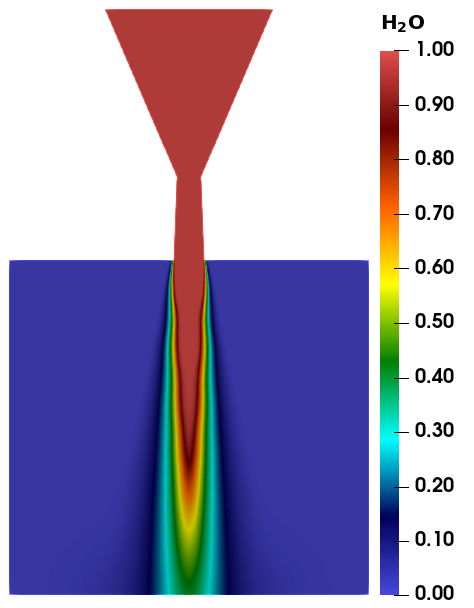}  
  \caption{}
  \label{fig:hconth}
\end{subfigure}

\vspace{1cm}
\hspace{3.6cm}
\begin{subfigure}{.5\textwidth}

  \centering
  \includegraphics[width=.8\linewidth,height=.7\linewidth]{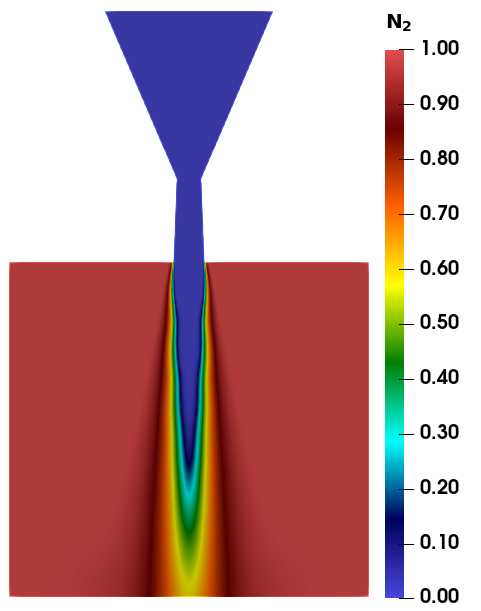}  
  \caption{}
  \label{fig:n2conth}
\end{subfigure}
\caption{The contours of (a) Pressure (b) Temperature (c) Mach number (d) $H_2O$ (e) $N_2$ for pure $H_2O$ plume}
\label{fig:contourh}
\end{figure}

\begin{figure}
\begin{subfigure}{.5\textwidth}
  \centering
  \includegraphics[width=.9\linewidth,height=.9\linewidth]{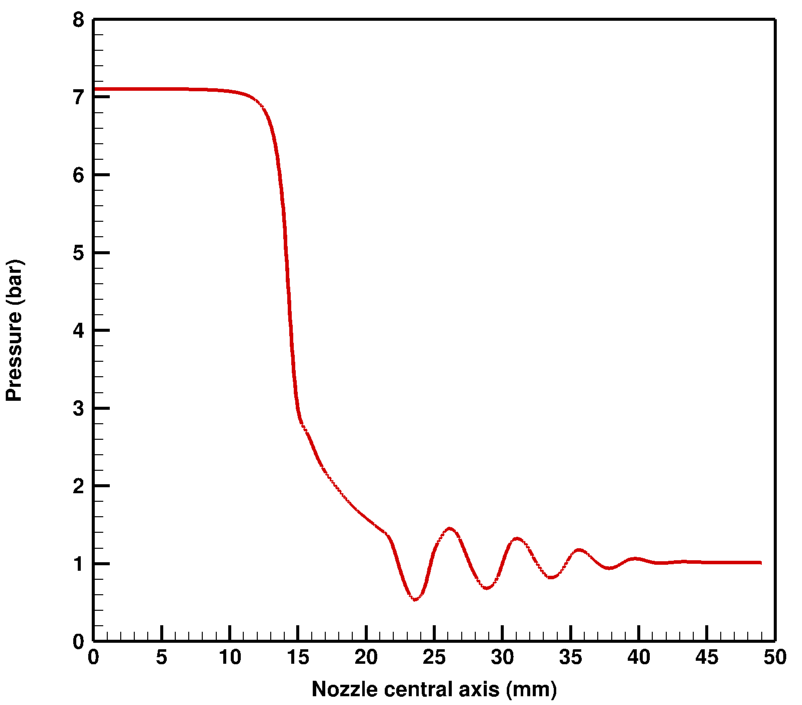}  
  \caption{}
  \label{fig:pcenth}
\end{subfigure}
\begin{subfigure}{.5\textwidth}
  \centering
  \includegraphics[width=.9\linewidth,height=.9\linewidth]{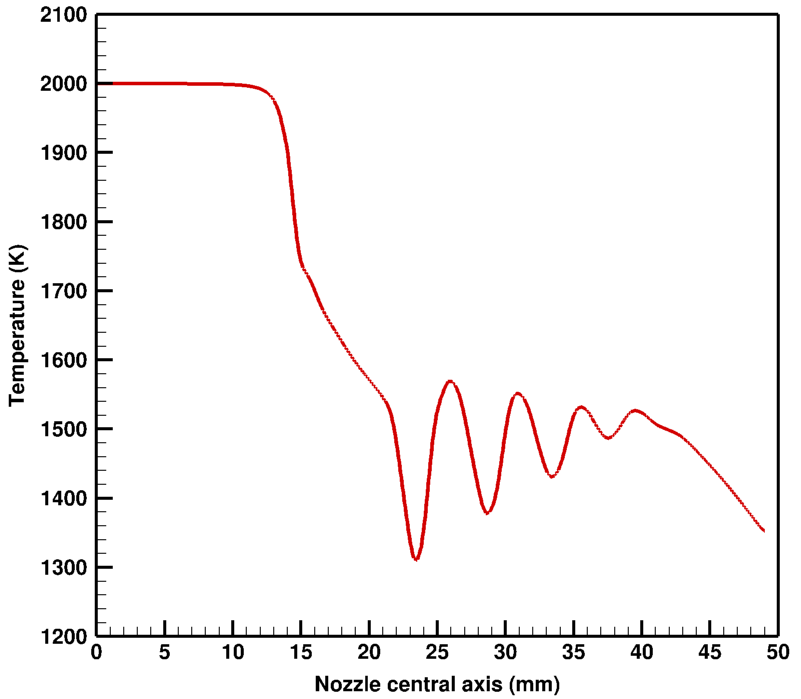}  
  \caption{}
  \label{fig:tcenth}
\end{subfigure}

\vspace{2cm}

\begin{subfigure}{.5\textwidth}
  \centering
  \includegraphics[width=.9\linewidth,height=.9\linewidth]{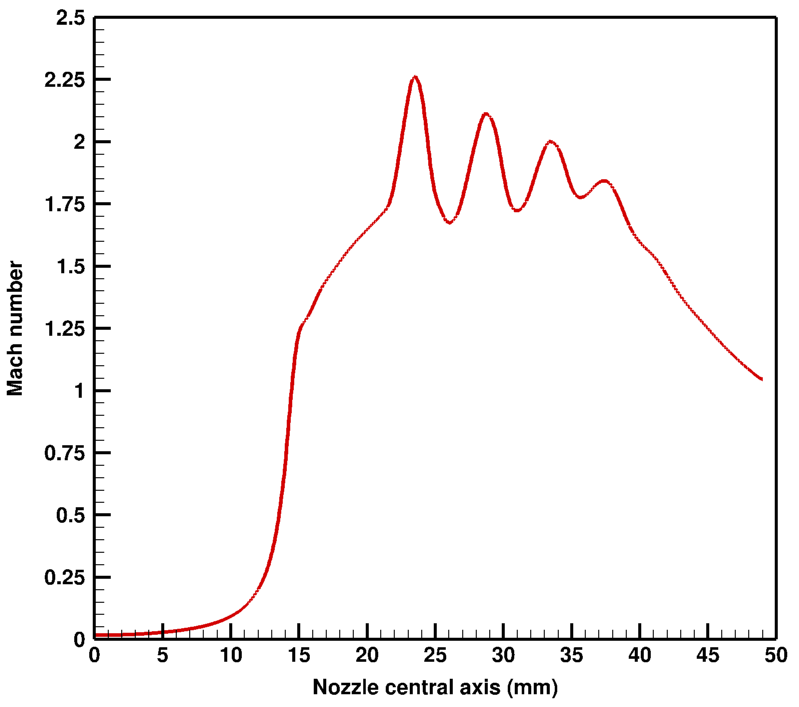}  
  \caption{}
  \label{fig:macenth}
\end{subfigure}
\begin{subfigure}{.5\textwidth}
  \centering
  \includegraphics[width=.9\linewidth,height=.9\linewidth]{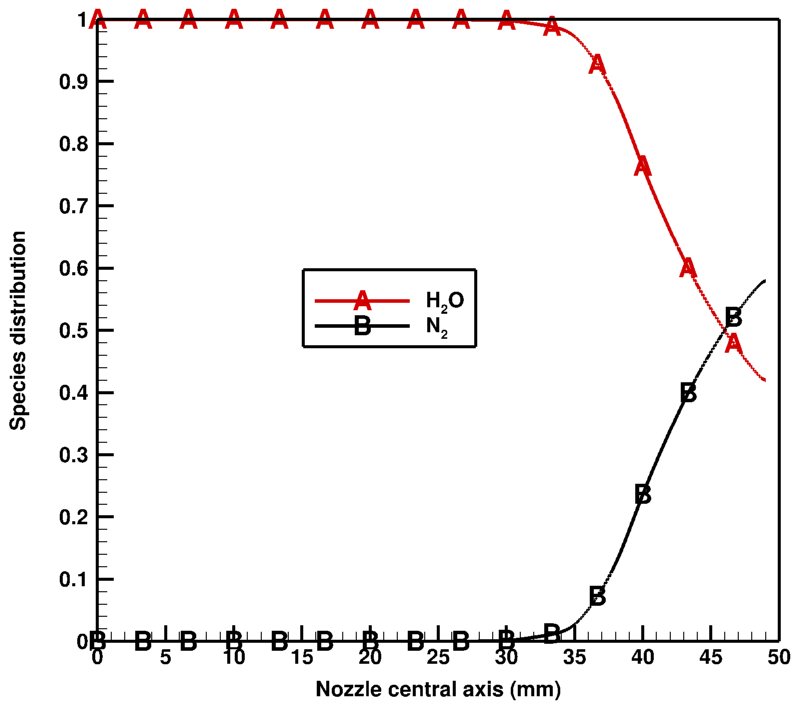}  
  \caption{}
  \label{fig:speciescenth}
\end{subfigure}
\caption{Profile of (a) Pressure (b) Temperature (c) Mach number (d) Species along the centerline for pure $H_2O$ plume}
\label{fig:centh}
\end{figure}

\begin{figure}
\begin{subfigure}{.5\textwidth}
  \centering
  \includegraphics[width=.8\linewidth,height=1\linewidth]{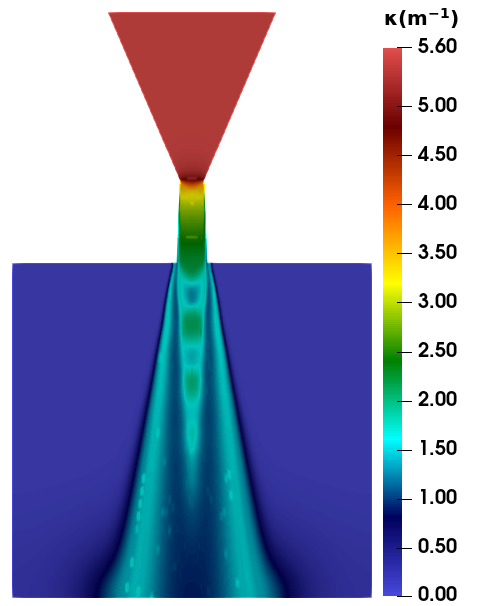}  
  \caption{}
  \label{fig:kconth}
\end{subfigure}
\begin{subfigure}{.5\textwidth}
  \centering
  \includegraphics[width=.8\linewidth,height=1\linewidth]{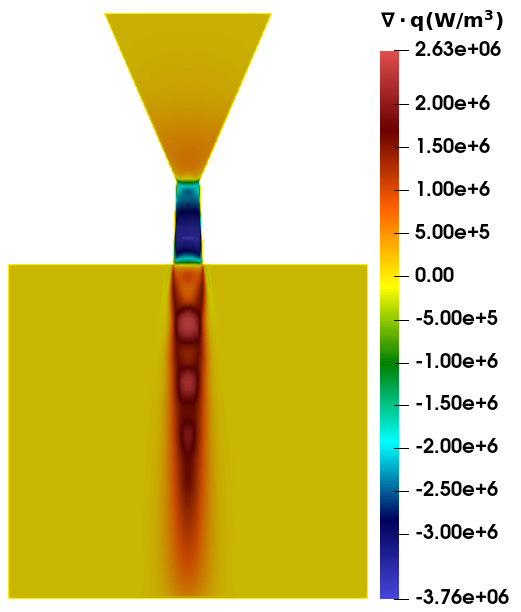}  
  \caption{}
  \label{fig:dqconth}
\end{subfigure}

\caption{The contours of (a) Absorption coefficient (b) Divergence of radiative heat flux for pure $H_2O$ plume}
\label{fig:contour2h}
\end{figure}

Pure $H_2O$ plume is formed by the combustion of pure $H_2$ with liquid oxidizer $LOX$. The resulting product contains mole-fraction of $H_2O$ ($x=1$) which emanates from the nozzle in the form of the plume. Initially the medium is filled with $N_2$, and the $H_2O$ expands from 7.11 bar and 2000 K to a quiescent medium of 1 atm and 288 K Temperature.\\
\indent The pressure remains constant in the convergent part of the nozzle, however it suddenly decreases at the throat and the divergent part of the nozzle as shown in Figure.\ref{fig:pcenth}.  The exit pressure at nozzle for $H_2O$ plume is slightly higher than the pressure of quiescent medium, i.e., 1.4 bar, this essentially means that the flow is underexpanded \cite{FRANQUET201525}. Due to this underexpansion scenario, there forms the series of strong expansion and compression waves (oblique shocks) which evolves from the lip of the nozzle, as pressure tries to adjust itself against medium pressure. The shock which evolve from the lip of the nozzle is in the shape of barrel so it can be called as "barrel shock" and a Mach disc appears after the shock which is formed due to singular reflection. The pressure variation in divergent part of the nozzle enables the temperature reduction as shown in Figure. 10b. Similar effect of pressure variation in the plume is seen on the temperature variation as well. Thus, the temperature variation in the divergent part of the nozzle and in the plume enables the heat transfer mechanism. However, heat transfer mechanism does not occur in the convergent part of the nozzle, due to uniform temperature inside the convergent part of the nozzle. The physical quantities such as pressure, temperature and velocity or Mach number vary rapidly across the shock. The shock pattern is in the form of a diamond also known as diamond flow structure. The pressure varies between 1.4 bar to 0.58 bar across the shock as in Figure 10a. Similarly, the temperature also varies sharply, i.e., upto 300 K in the region from 23 mm to 25 mm as it can be seen from temperature profile across the axis in Fig. \ref{fig:tcenth}. The temperature first decreases due to expansion of gases and then it increases due to compression wave and this pattern continues till pressure comes in equilibrium with the buffer zone pressure. After 40 mm, flow stabilizes, as the pressure of fluid at that point becomes same as that of medium pressure. The trend is opposite for Mach number as gas expands, the velocity of the flow increases and the maximum value of Mach number achieved in this case is 2.25. The contour of Mach number and its profile along the centerline distribution are shown in Fig. \ref{fig:maconth} and \ref{fig:macenth}, respectively. In the near field region of plume, after the inviscid core central region, there forms a mixing layer where viscosity effects are felt and the primary species ($H_2O$) starts getting mixed with the atmospheric species ($N_2$) and forms shear layer. The region just outside the nozzle where species starts mixing is called as entrainment region of the plume. Moving downstream in the direction of the flow, mixing layer widens for $H_2O$ being lighter molecule (molecular weight=18), as in Fig. \ref{fig:hconth}. In the far field region, i.e., the region after the shock, species mixes completely till the centerline as it can be seen in the $H_2O$ and $N_2$ profiles along the centerline. Fig. \ref{fig:speciescenth} shows the profiles of $H_2O$ and $N_2$ along the axis and contours of $H_2O$ and $N_2$ are represented in the Figs. \ref{fig:hconth} and \ref{fig:n2conth}, respectively. \\
\indent The pressure, temperature and the species concentration of $H_2O$ contours constitutes the thermodynamic state of the $H_2O$ vapour, and Planck mean absorption coefficient of $H_2O$ has been accessed through lookup tables and its contours is shown in Figure.\ref{fig:kconth}. It has very high value in the convergent portion of the nozzle due to very high pressure and decreases as pressure decreases in the divergent section of the nozzle and its value is further reduced in the plume. The absorption coefficient is zero where only $N_2$ gas is available plume being very small thickness, the reabsorption does not occur and the major emission comes from the core of the plume, as emission and absorption are almost same in the shear layer as the divergence of radiative heat flux is almost zero in the shear layer and the regions of zero absorption coefficient as shown in Figure. 11b. One thing to notice that the range of divergence of radiative flux is negative to positive, both the positive value of the divergence of radiative flux reveals radiative sink term while negative value tells radiative source term, Thus, radiation is heating the gas inside the divergent part of the nozzle while it is cooling the plume. Further the energy is transferred by radiation mode of heat transfer to other region without any change.

\begin{figure}
\begin{subfigure}{.5\textwidth}
  \centering
  \includegraphics[width=.9\linewidth,height=.9\linewidth]{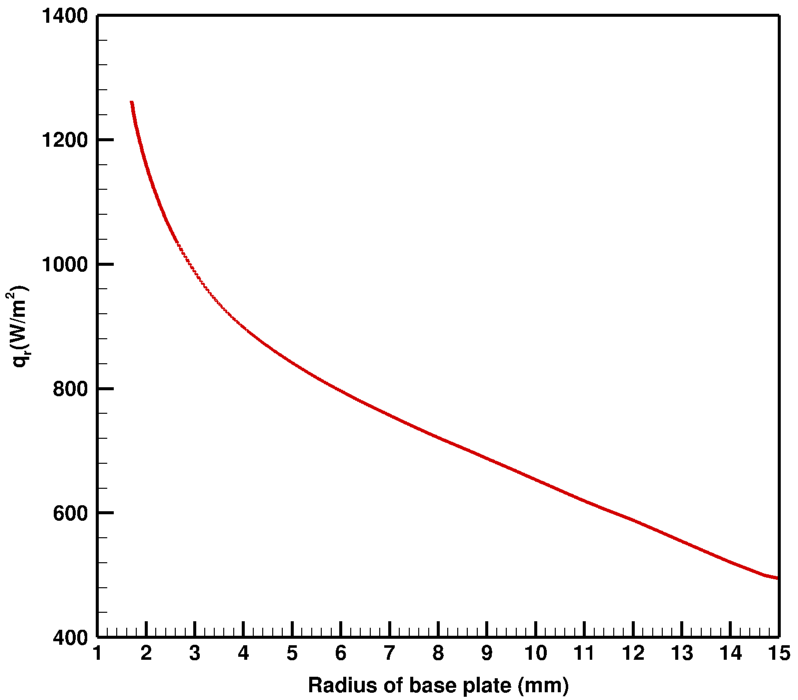}  
  \caption{}
  \label{fig:qrlath}
\end{subfigure}
\begin{subfigure}{.5\textwidth}
  \centering
  \includegraphics[width=.9\linewidth,height=.9\linewidth]{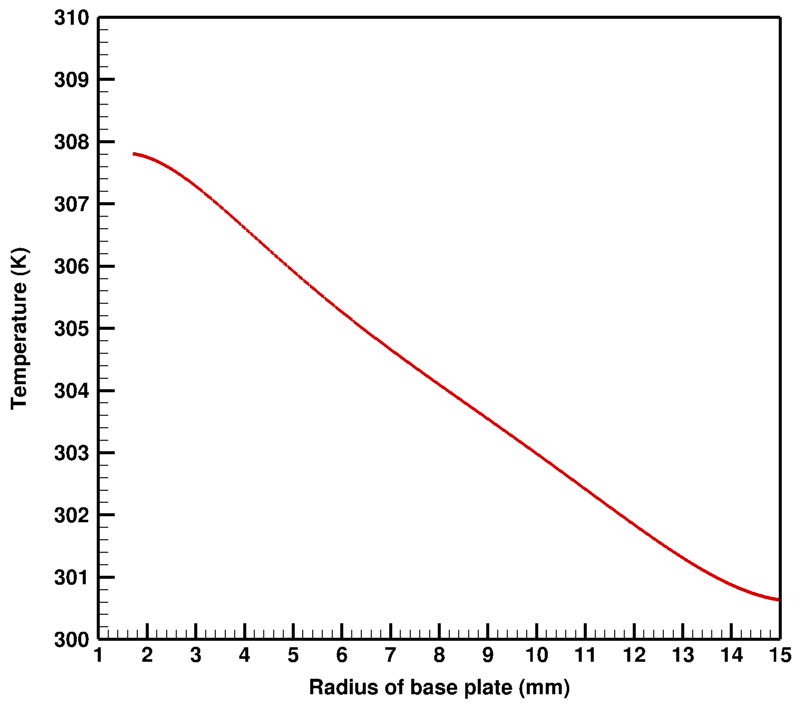}  
  \caption{}
  \label{fig:tlath}
\end{subfigure}

\caption{Profile of (a) Radiative heat flux (b) Temperature along the radius of the base plate for pure $H_2O$ plume}
\label{fig:lateralh}
\end{figure}
\indent The high temperature plume after emanating from the nozzle gets diffused and develop very high flux and temperature in a very narrow region around the lip of the nozzle on the base plate. Barring this region, the base plate receives the radiation energy emanating from the shear layer of plume. The radiative heat flux on the base plate is shown in Fig. \ref{fig:qrlath}, baring some region near to the lip of the nozzle. The maximum value of radiative heat flux is 1300 $W/m^2$ and it decreases along the radial direction as the view factor of plume decreases. Similarly, the temperature developed due to this radiative flux is shown in Fig. \ref{fig:tlath}. The maximum value which base plate attains due to radiation energy is 308 K and it decreases in the similar manner of radiation flux along the radius.

\subsection{Pure $CO_2$ plume}
\begin{figure}
\begin{subfigure}{.5\textwidth}
  \centering
  \includegraphics[width=.8\linewidth,height=.7\linewidth]{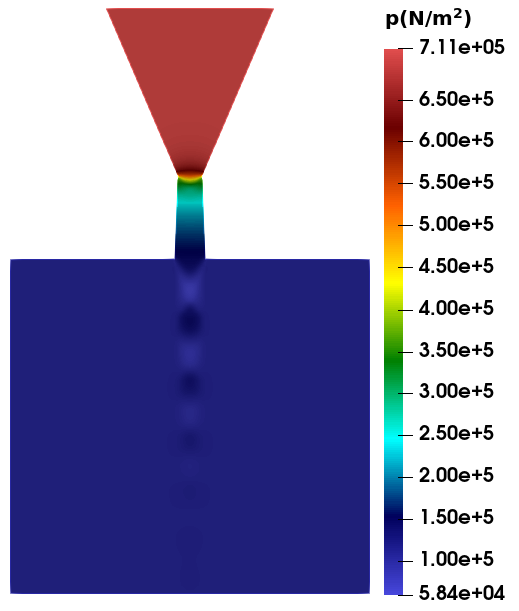}  
  \caption{}
  \label{fig:pcontc}
\end{subfigure}
\begin{subfigure}{.5\textwidth}
  \centering
  \includegraphics[width=.8\linewidth,height=.7\linewidth]{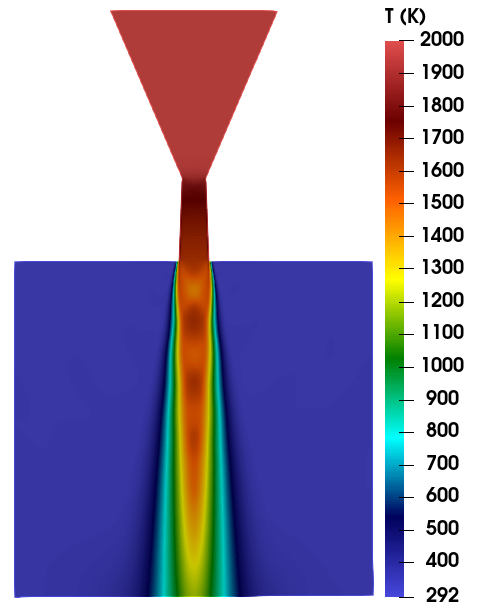}  
  \caption{}
  \label{fig:tcontc}
\end{subfigure}

\vspace{1cm}

\begin{subfigure}{.5\textwidth}
  \centering
  \includegraphics[width=.8\linewidth,height=.7\linewidth]{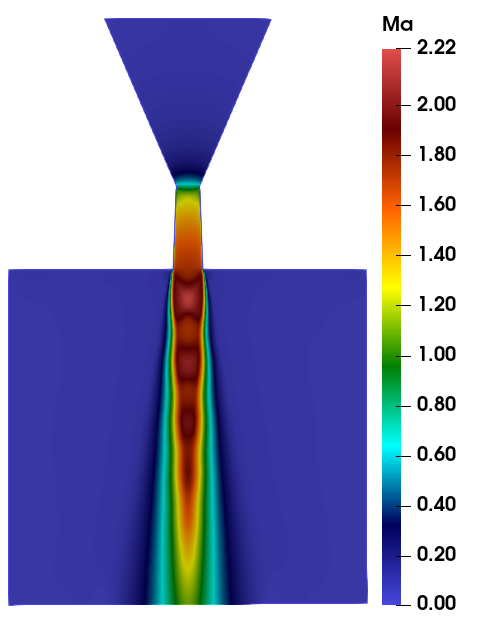}  
  \caption{}
  \label{fig:macontc}
\end{subfigure}
\begin{subfigure}{.5\textwidth}
  \centering
  \includegraphics[width=.8\linewidth,height=.7\linewidth]{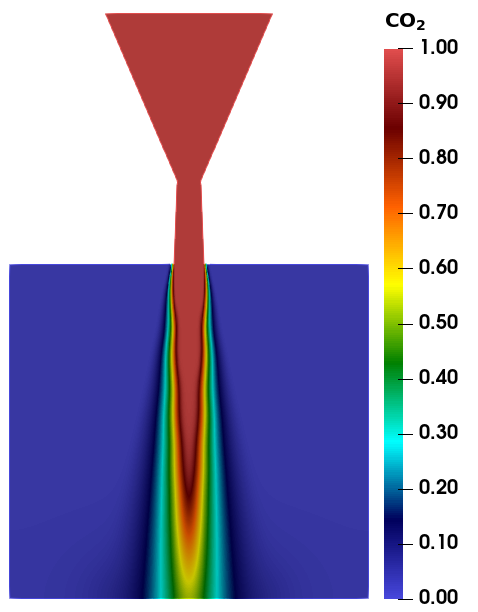}  
  \caption{}
  \label{fig:ccontc}
\end{subfigure}

\vspace{1cm}

\hspace{3.6cm}
\begin{subfigure}{.5\textwidth}
  \centering
  \includegraphics[width=.8\linewidth,height=.7\linewidth]{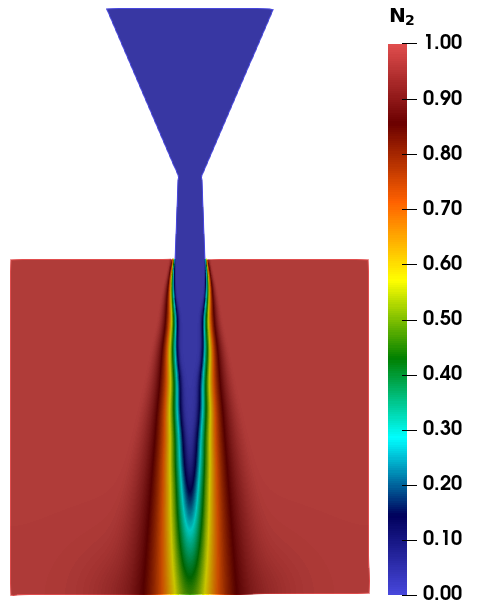}  
  \caption{}
  \label{fig:n2contc}
\end{subfigure}
\caption{Contours of (a) Pressure (b) Temperature (c) Mach number (d) $CO_2$ (e) $N_2$ for pure $CO_2$ plume}
\label{fig:contourc}
\end{figure}

\begin{figure}
\begin{subfigure}{.5\textwidth}
  \centering
  \includegraphics[width=.9\linewidth,height=.9\linewidth]{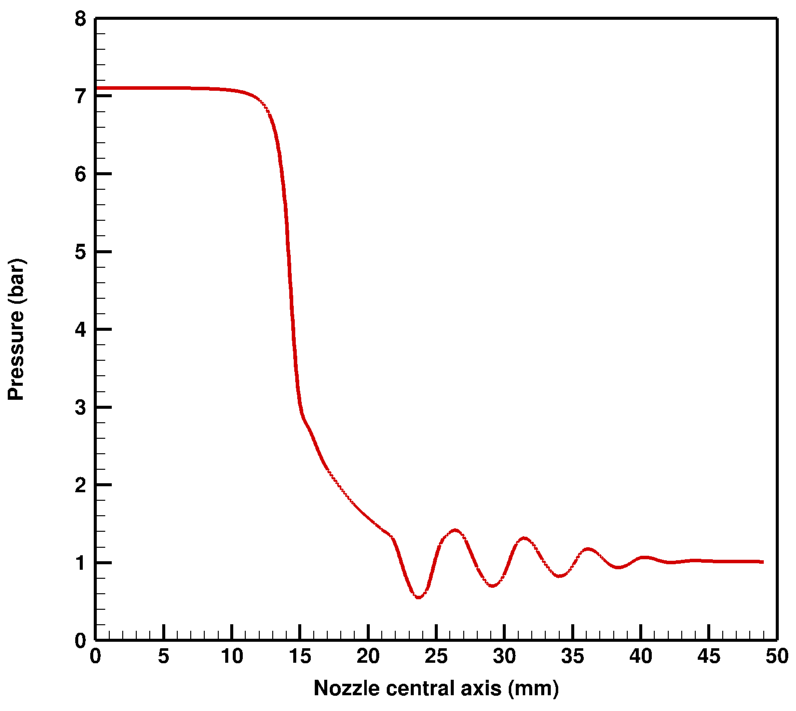}  
  \caption{}
  \label{fig:pcentc}
\end{subfigure}
\begin{subfigure}{.5\textwidth}
  \centering
  \includegraphics[width=.9\linewidth,height=.9\linewidth]{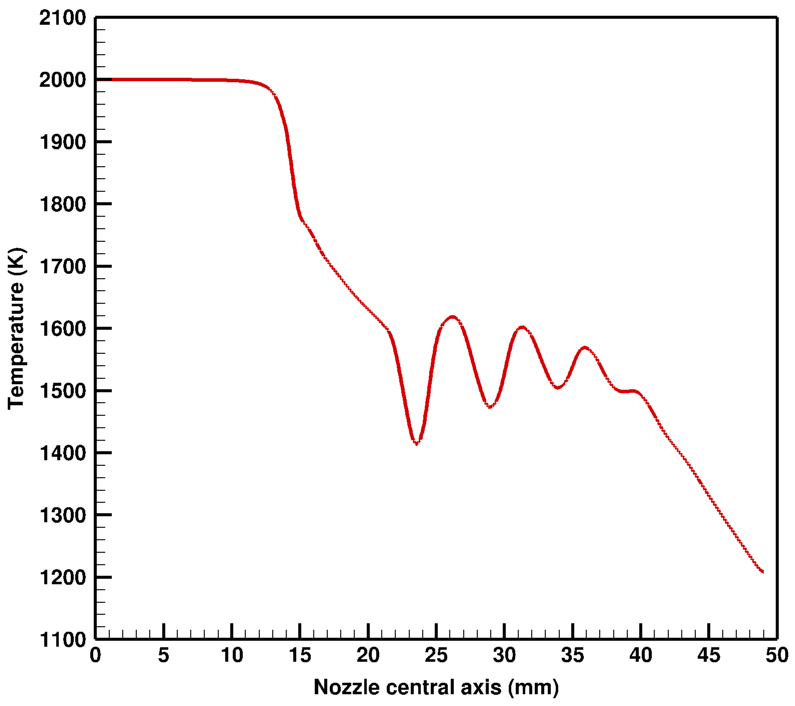}  
  \caption{}
  \label{fig:tcentc}
\end{subfigure}

\vspace{2cm}

\begin{subfigure}{.5\textwidth}
  \centering
  \includegraphics[width=.9\linewidth,height=.9\linewidth]{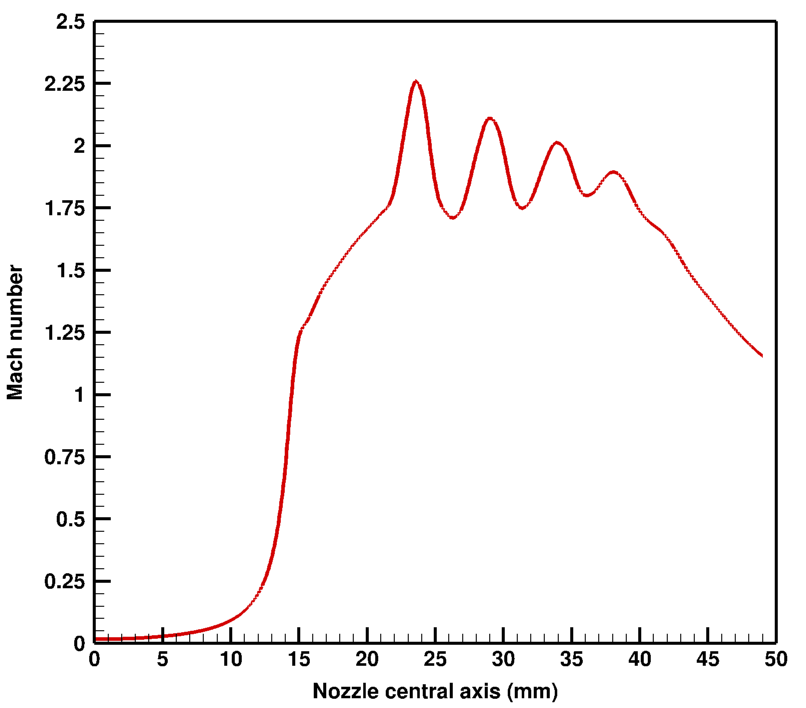}  
  \caption{}
  \label{fig:macentc}
\end{subfigure}
\begin{subfigure}{.5\textwidth}
  \centering
  \includegraphics[width=.9\linewidth,height=.9\linewidth]{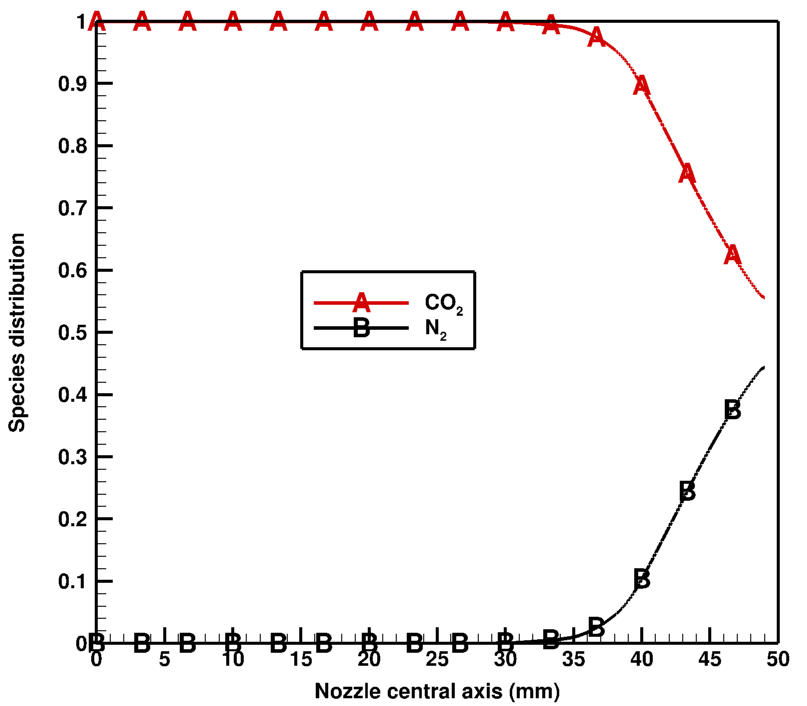}  
  \caption{}
  \label{fig:speciescentc}
\end{subfigure}
\caption{Profile of (a) Pressure (b) Temperature (c) Mach number (d) Species for pure $CO_2$ plume}
\label{fig:centc}
\end{figure}

 \indent Although generation of pure $CO_2$ plume is not very much realistic, however, for the theoretical understanding the simulation has been performed for pure $CO_2$ plume. The simulations for pure $CO_2$ are performed by supplying pure $CO_2$ ($x=1$) at the inlet of the nozzle and rest conditions are kept same as that of $H_2O$ plume. This is also the case of underexpansion, so pressure at the lip of the nozzle varies from 1.4 bar to 0.5 bar across the shocks. There is a formation of Mach disc at the end of the first shock. The contour of pressure and distribution of pressure along the centerline is shown in Fig. \ref{fig:pcontc} and \ref{fig:pcentc}, respectively. The temperature drop across the shock in the $CO_2$ plume is less compared to $H_2O$, however there is more drop in temperature towards the end of the plume.

\begin{figure}
\begin{subfigure}{.5\textwidth}
  \centering
  \includegraphics[width=.8\linewidth,height=1\linewidth]{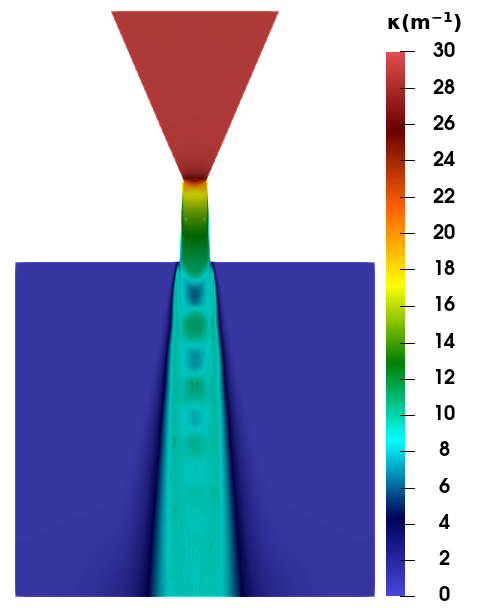}  
  \caption{}
  \label{fig:kcontc}
\end{subfigure}
\begin{subfigure}{.5\textwidth}
  \centering
  \includegraphics[width=.8\linewidth,height=1\linewidth]{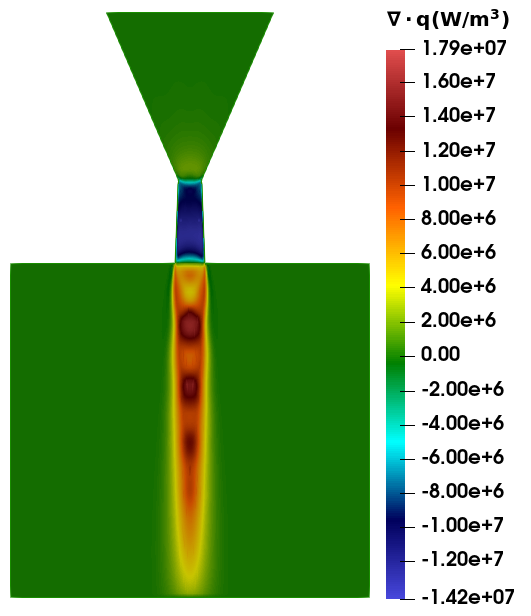}  
  \caption{}
  \label{fig:dqcontc}
\end{subfigure}
\caption{Contours of (a) Absorption coefficient (b) Divergence of radiative heat flux for pure $CO_2$ plume }
\label{fig:contour2h}
\end{figure}

\begin{figure}
\begin{subfigure}{.5\textwidth}
  \centering
  \includegraphics[width=.9\linewidth,height=.9\linewidth]{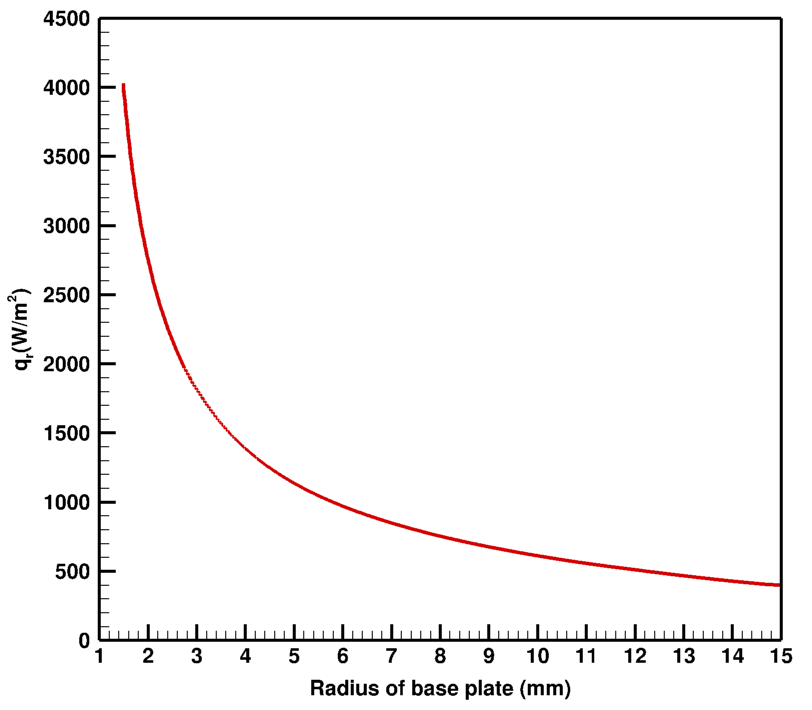}  
  \caption{}
  \label{fig:qrlatc}
\end{subfigure}
\begin{subfigure}{.5\textwidth}
  \centering
  \includegraphics[width=.9\linewidth,height=.9\linewidth]{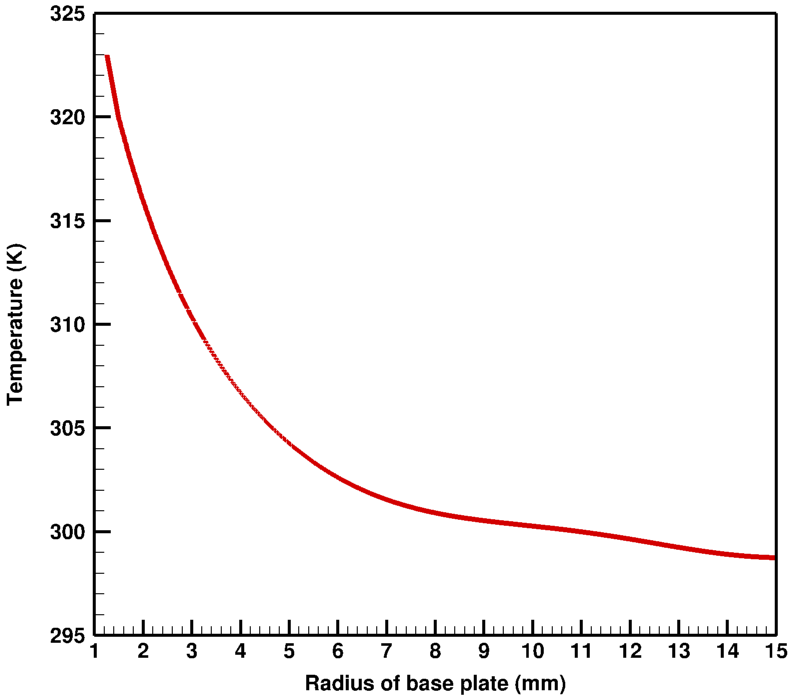}  
  \caption{}
  \label{fig:tlatc}
\end{subfigure}

\caption{Profile of (a) Radiative heat flux (b) Temperature along the radius of base plate for pure $CO_2$ plume}
\label{fig:lateralc}
\end{figure}
\noindent The temperature contour is shown in fig. \ref{fig:tcontc}. The variation of temperature across the first shock is not much drastic in comparison to $H_2O$ plume, also this plume cools faster than $H_2O$ plume i.e., minimum temperature of this plume is 1200 K at the ends while it is 1350 K for $H_2O$ plume. The Mach number contour and its distribution along the centerline are shown in Fig. \ref{fig:macontc} and \ref{fig:macentc}, respectively. The diffusion of $CO_2$ in $N_2$ is less in comparison to $H_2O$ due to higher molecular weight of $CO_2$ (44) compared to $H_2O$ (18) as shown in Fig. \ref{fig:speciescentc}. The contours of $CO_2$ and $N_2$ mole fraction are shown in Fig. \ref{fig:ccontc} and \ref{fig:n2contc}, respectively.\\
\indent The absorption coefficient distribution by considering Planck mean absorption coefficient for $CO_2$ plume is shown in Fig. \ref{fig:kcontc}. Its value is almost zero everywhere except in the core of the plume and in the shear layer. As the absorption coefficient of $CO_2$ is higher in the shear layer compared to $H_2O$ plume, the radiative heat flux on the rocket base plate is also higher, i.e., around 4000 $W/m^2$ as shown in Fig. \ref{fig:qrlatc}. The corresponding temperature distribution on the base plate is shown in Fig. \ref{fig:tlatc}, having a maximum value of 323 K, barring the diffusion region.

\begin{figure}
\begin{subfigure}{.49\textwidth}
  \centering
  \includegraphics[width=.9\linewidth,height=.9\linewidth]{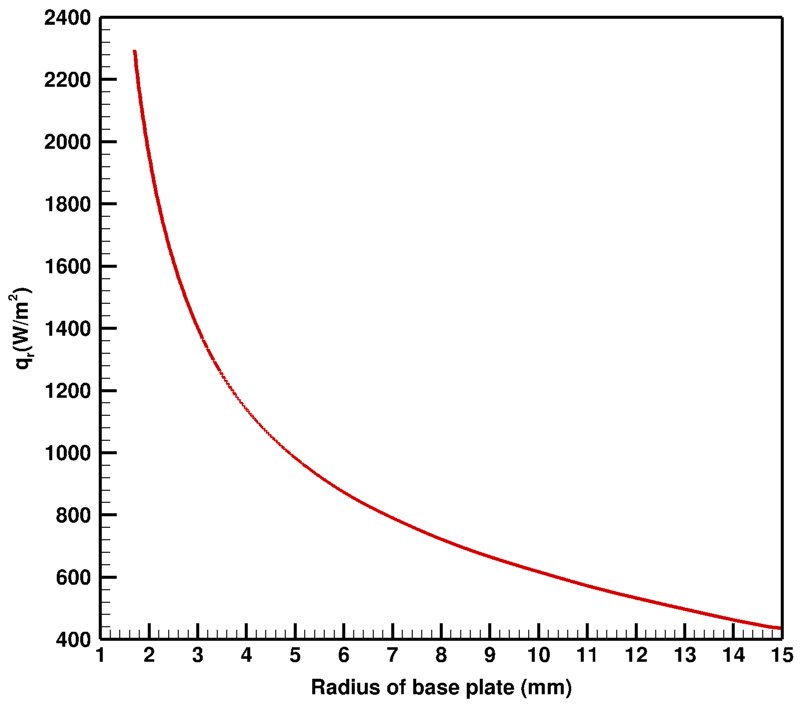}  
  \caption{}
  \label{fig:qrlatm}
\end{subfigure}
\begin{subfigure}{.49\textwidth}
  \centering
  \includegraphics[width=.9\linewidth,height=.9\linewidth]{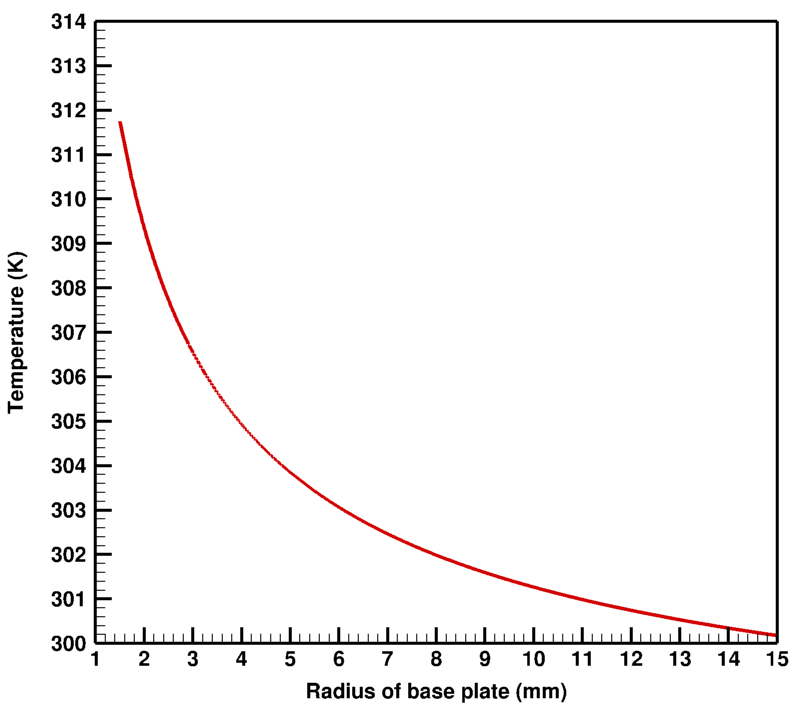}  
  \caption{}
  \label{fig:tlatm}
\end{subfigure}

\caption{Profile of (a) Radiative heat flux (b) Temperature on base plate along the radius of base plate for 50-50\% mixture of $CO_2-H_2O$ plume }
\label{fig:lateralm}
\end{figure}
\subsection{Mixture plume ($ 50 \;\% \;H_2O$ and $50\; \%\; CO_2$)}
\indent The combustion of hydrocarbon fuel with liquid oxidizer ($LOX$) gives 50-50\% mixture of $CO_2$ and $H_2O$. Thus, for the present problem, we supply 50-50\% mixture of both $CO_2$ and $H_2O$ at the inlet of the nozzle and other conditions are kept same as previous cases for the simulation of this plume. This is also a case of underexpanded plume. The temperature variation along the centerline at the end of the buffer section is somewhat the average of both pure $CO_2$ and $H_2O$ plume. \\
\indent The radiative transfer calculations are performed to determine the heat flux on the base plate from $CO_2-H_2O$ plume. The maximum radiative heat flux on the base plate is 2300 $W/m^2$ (Fig. \ref{fig:qrlatm}) and it decays with the radius of the base plate. The corresponding profiles of the temperature on the base plate is shown in Fig. \ref{fig:tlatm}. It is noted that the flux and temperature profiles for $CO_2$ and mixture plume are exponential decaying with radius, while it is almost linear for $H_2O$. This is owing to the fact that, high diffusion of $H_2O$ causes more spreading of $H_2O$ and this emission from $H_2O$ has high view factor, while this is not the case with $CO_2$ and mixture plume. 

\section{\textbf{Conclusions}}
The thermal load calculation on the base plate of nozzle from exhaust plume is performed in OpenFOAM. The ability of pressure based compressible flow application, "sonicFoam" is tested to capture the flow fields for air expanding in a convergent divergent nozzle. The stagnation pressure and temperature at the inlet of the nozzle are 7.11 bar and 288 K due to which flow expands and achieves Mach 2.1 at the exit of the nozzle. The resulting pressure and Mach number variation at the centerline matches well with the standard published results.\\ 
\indent The same nozzle is then used with elevated stagnation temperature of 2000 K and same pressure at inlet, to estimate the heat load on base plate for three different plumes namely, pure $H_2O$ plume, pure $CO_2$ plume and mixture plume. The "sonicFoam" application is then modified by incorporating the work done due to viscous forces, species transport equation and finally clubbed with the RTE solver fvDOM along with Planck mean absorption emission model and named as "radSonicFOAM". 
All three plumes exit from the nozzle at underexpanded flow conditions, where exit pressure is higher than the back pressure. The expansion waves start from the lip of the nozzle due to which the temperature decreases as flow exit from the nozzle and Mach number increases to a maximum value of 2.25.\\
\indent The maximum amount of heat load in the present study due to thermal radiation on base plate is from pure $CO_2$ plume, i.e., 4000 $W/m^2$ due to the high value of absorption coefficient, barring the diffusion zone. This flux heats up the base plate and its temperature rises upto 323 K, followed by mixture plume, which receives maximum radiative heat flux of 2300 $W/m^2$ and the corresponding rise in temperature is 312 K. For pure $H_2O$ plume, the heat flux is least, i.e., 1300 $W/m^2$ with temperature rise of 308 K. For different plumes the variation in flux is different and this is mostly due to the difference in the absorption coefficient of the gases. Further, their molecular weights are also different, due to which there is difference in the flow field of the gases and also the different nature of flux and temperature variations on the nozzle base plate.

Due to small length scale, the current case falls in optically thin regime, thus, the Planck mean absorption model provides the satisfactory results, however, Planck mean absorption model may not be useful for other cases with big length scale. Therefore, The full spectrum radiative properties models are needed with the properties for all thermodynamic states existing in the plume. Furthermore, the solid fuel emanates particles which contribute most of the radiative thermal load on the nozzle base plate, therefore the current radiation heat transfer feature needs to further enhance by including the scattering model. 
\bibliography{mybibfile}
\end{document}